\documentclass[useAMS,usenatbib]{mn2e}

\usepackage[letterpaper,margin=0.635in]{geometry}
\usepackage[ansinew]{inputenc}
\usepackage{amsfonts}
\usepackage{amsmath}
\usepackage{enumerate}
\usepackage{enumitem}
\usepackage{graphicx}
\usepackage{epsfig}
\usepackage{color}
\usepackage[normalem]{ulem}
\usepackage{comment}
\usepackage{amssymb}
\usepackage{pdflscape}
\usepackage{float}

\usepackage[T1]{fontenc}
\usepackage{aecompl}

\usepackage[pagewise]{lineno}

\def\redmagic{redMaGiC}

\newcommand{\pz}{photo-$z$\ }
\newcommand{\pzs}{photo-$z$'s\ }

\newcommand{\mpch}{{Mpc/$h$}}

\newcommand{\zl}{z_{\rm L}}
\newcommand{\zs}{z_{\rm s}}

\newcommand{\be}{\begin{equation}}
\newcommand{\ee}{\end{equation}}
\newcommand{\bes}{\begin{equation*}}
\newcommand{\ees}{\end{equation*}}
\newcommand{\bea}{\begin{eqnarray}}
\newcommand{\eea}{\end{eqnarray}}
\newcommand{\beas}{\begin{eqnarray*}}
\newcommand{\eeas}{\end{eqnarray*}}

\newcommand{\cs}[1]{{\textcolor{black}{#1}}}

\newcommand{\photoz}{photo-$z$}

\title[Cosmic Voids and Void Lensing in the DES-SV data]{Cosmic Voids and Void Lensing in the Dark Energy Survey Science Verification Data}

\author[S{\'a}nchez et al.]{
\parbox{\textwidth}{
\Large
C.~S\'{a}nchez$^1$\footnotemark, 
J.~Clampitt$^{2}$,
A.~Kovacs$^{1}$,
B.~Jain$^{2}$,
J.~Garc\'ia-Bellido$^{3}$,
S.~Nadathur$^{4}$,
D.~Gruen$^{5,6,7}$,
N.~Hamaus$^{8}$,
D.~Huterer$^{9}$,
P.~Vielzeuf$^{1}$,
A.~Amara$^{10}$,
C.~Bonnett$^{1}$,
J.~DeRose$^{11,5}$,
W.~G.~Hartley$^{12,10}$,
M.~Jarvis$^{2}$,
O.~Lahav$^{12}$,
R.~Miquel$^{13,1}$,
E.~Rozo$^{14}$,
E.~S.~Rykoff$^{5,6}$,
E.~Sheldon$^{15}$,
R.~H.~Wechsler$^{11,5,6}$,
J.~Zuntz$^{16}$,
T. M. C.~Abbott$^{17}$,
F.~B.~Abdalla$^{12,18}$,
J.~Annis$^{19}$,
A.~Benoit-L{\'e}vy$^{20,12,21}$,
G.~M.~Bernstein$^{2}$,
R.~A.~Bernstein$^{22}$,
E.~Bertin$^{20,21}$,
D.~Brooks$^{12}$,
E.~Buckley-Geer$^{19}$,
A. Carnero Rosell$^{23,24}$,
M.~Carrasco~Kind$^{25,26}$,
J.~Carretero$^{27,1}$,
M.~Crocce$^{27}$,
C.~E.~Cunha$^{5}$,
C.~B.~D'Andrea$^{4,28}$,
L.~N.~da Costa$^{23,24}$,
S.~Desai$^{29,30}$,
H.~T.~Diehl$^{19}$,
J.~P.~Dietrich$^{29,30}$,
P.~Doel$^{12}$,
A.~E.~Evrard$^{31,9}$,
A.~Fausti Neto$^{23}$,
B.~Flaugher$^{19}$,
P.~Fosalba$^{27}$,
J.~Frieman$^{19,32}$,
E.~Gaztanaga$^{27}$,
R.~A.~Gruendl$^{25,26}$,
G.~Gutierrez$^{19}$,
K.~Honscheid$^{33,34}$,
D.~J.~James$^{17}$,
E.~Krause$^{5}$,
K.~Kuehn$^{35}$,
M.~Lima$^{36,23}$,
M.~A.~G.~Maia$^{23,24}$,
J.~L.~Marshall$^{37}$,
P.~Melchior$^{38}$,
A.~A.~Plazas$^{39}$,
K.~Reil$^{6}$,
A.~K.~Romer$^{40}$,
E.~Sanchez$^{3}$,
M.~Schubnell$^{9}$,
I.~Sevilla-Noarbe$^{3}$,
R.~C.~Smith$^{17}$,
M.~Soares-Santos$^{19}$,
F.~Sobreira$^{41,23}$,
E.~Suchyta$^{42}$,
G.~Tarle$^{9}$,
D.~Thomas$^{4}$,
A.~R.~Walker$^{17}$,
J.~Weller$^{29,43,8}$}
  \vspace{0.4cm}\\~\\
\parbox{\textwidth}{\centering \textsc{\Large(The DES Collaboration)} \\ \centering \textit{Author affiliations are listed at the end of this paper} }
\vspace{-1cm}
}
\begin{document}



\pagerange{\pageref{firstpage}--\pageref{lastpage}} \pubyear{0000}
\maketitle
\label{firstpage}
\begin{abstract}
Galaxies and their dark matter halos populate a complicated filamentary network around large, nearly empty regions known as cosmic voids. Cosmic voids are usually identified in spectroscopic galaxy surveys, where 3D information about the large-scale structure of the Universe is available.
Although an increasing amount of photometric data is being produced, its potential for void studies is limited since photometric redshifts induce line-of-sight position errors of $\sim50$ Mpc/$h$ or more that can render many voids undetectable.
In this paper we present a new void finder designed for photometric surveys, validate it using simulations, and apply it to the high-quality photo-$z$ \redmagic{} galaxy sample of the Dark Energy Survey Science Verification (DES-SV) data.
The algorithm works by projecting galaxies into 2D slices and finding voids in the smoothed 2D galaxy density field of the slice.
Fixing the line-of-sight size of the slices to be at least twice the photo-$z$ scatter, the number of voids found in \cs{these projected slices of} simulated spectroscopic and photometric galaxy catalogs is within 20\% for all transverse void sizes, and indistinguishable for the largest voids of radius $\sim 70$ \mpch~and larger.
The positions, radii, and projected galaxy profiles of photometric voids also accurately match the spectroscopic void sample.
Applying the algorithm to the DES-SV data \cs{in the redshift range $0.2<z<0.8$, we identify 87 voids with comoving radii spanning the range 18-120 Mpc/$h$}, and carry out a stacked weak lensing measurement.
With a significance of $4.4\sigma$, the lensing measurement confirms the voids are truly underdense in the matter field and hence not a product of Poisson noise, tracer density effects or systematics in the data. It also demonstrates, for the first time in real data, the viability of void lensing studies in photometric surveys. 
\end{abstract}

\begin{keywords}
large-scale structure of Universe -- cosmology: observations -- gravitational lensing: weak 
\end{keywords}

\section{Introduction}
\label{sec:intro}
\renewcommand*{\thefootnote}{\fnsymbol{footnote}}
\footnotetext[1]{
Corresponding author: \texttt{csanchez@ifae.es}}

Cosmic voids are low-density regions in space surrounded by a network of dark matter halos and the galaxies that populate them. Given their intrinsic low-density environment, voids are only weakly affected by complicated non-linear gravitational effects which have a strong impact in crowded environments such as galaxy clusters. This simplicity makes it possible to constrain cosmological parameters with voids \citep{Betancort-Rijo2009,Lavaux2010,Sutter2014b,Kitaura2015,Hamaus2016,Mao2016,Sahlen2016}. Furthermore, the unique low-density environments of voids make possible probes of the nature of dark energy, alternate theories of gravity \citep{Lee2009,Bos2012,Spolyar2013,Cai2015,Barreira2015}, and primordial non-Gaussianity \citep{Song2009}. 

A number of different void finding algorithms exist in the literature: Voronoi tesselation and watershed methods \citep{Platen2007, Neyrinck2008,Lavaux2012,Sutter2012,Nadathur2015}, growth of spherical underdensities \citep{Hoyle2002, Colberg2005, Padilla2005, Ceccarelli2006, Li2011}, hybrid methods \citep{Jennings2013}, 2D projections \citep{Clampitt2014}, dynamical criteria \citep{Elyiv2015}, and Delaunay Triangulation \citep{Zhao2015}, among other methods \citep{Colberg2008}.
Most void finders currently applied to data use galaxies with spectroscopic redshifts to define voids. However, when using far less precise photometric redshifts (photo-$z$'s), the void-finding process needs to be revisited to overcome the smearing in the line-of-sight position of tracer galaxies.

Spectroscopic surveys like 2dF \citep{Colless2001}, VVDS \citep{LeFevre2005}, WiggleZ \citep{Drinkwater2010} or BOSS \citep{Dawson2013} provide 3D information of the galaxy distribution, but they are expensive in terms of time, and may suffer from selection effects, incompleteness and limited depth. In contrast, photometric surveys such as SDSS \citep{York:2000gk}, PanSTARRS \citep{Kaiser2000}, KiDS \citep{Jong2013} or LSST \citep{Tyson:2002nh} are more efficient and nearly unaffected by selection bias, more complete and deeper, but do not provide complete 3D information of the galaxy distribution due to their limited resolution in the galaxy line-of-sight positions, obtained by measuring the photo-$z$ of each galaxy from the fluxes measured through a set of broadband filters. 

A few void catalogs exist that use photometric redshift tracers \citep{Granett2008}. Many voids about the size of the \pz error or smaller will not be found at all; in other cases, spurious, or \textit{Poisson}, voids will appear in the sample due to \pz scatter. For the larger voids in the sample, those with sizes much larger than the \pz error, the \pz scatter should not affect the void sample substantially. However, these huge voids are very few due to the rapidly falling size distribution of cosmic voids in the universe. \cs{In any case, it should also be possible to find voids smaller than the photo-$z$ scatter, since the latter acts to smooth out the density field, but retains the topology of the large-scale structure to some extent.} Therefore, by designing a void finding algorithm specifically for photometric redshift surveys, the purity and completeness of the resulting void sample can be improved.

Qualitatively, our void finding method can be understood with an analogy to galaxy clustering measurements.
In that case, the ideal scenario is to measure the 3D correlation function of galaxies when spectroscopic redshifts are available.
However, for photometric survey data sets, one \cs{usually avoids computing} the 3D correlation function of galaxies because of the \pz dispersion affecting the line-of-sight component.
The standard approach is therefore to split galaxies into tomographic photometric redshift bins, and compute the 2D angular correlation function in \cs{the projection of} each of these line-of-sight bins.
The photometric redshift errors make the actual size of the redshift bins to be effectively comparable or larger than the \pz scatter (see for instance \citealt{Crocce2011}). Then, in order to minimize the noise in the measurement, the optimal approach is to set the width of the redshift bins to be comparable or larger than the photo-$z$ scatter. Finally, one measures the angular clustering in each of these redshift bins, and hence the evolution of clustering with redshift.
In this work we present a void finder that follows the same approach: finding voids in the angular projection of the galaxy distribution in redshift slices that are broader than the \pz dispersion, and then combining the slices to get the most of the line-of-sight information in the data.

Before applying the algorithm to the DES Science Verification (DES-SV) data set, we use simulations with mock spectroscopic and realistic photometric redshifts to validate the method\cs{, running the void finder in both cases and studying the differences among the void catalogs coming from the corresponding projected slices}. Once the DES-SV void catalog is defined, we measure the weak gravitational lensing signal around voids and confirm the voids are also empty in the dark matter.

The plan of the paper is as follows. In Sec.~2 we describe the Dark Energy Survey Science Verification data used in this paper, together with the simulations used to test the validity of the finder. Section 3 presents the 2D angular void finder algorithm and some simulation tests comparing the algorithm output when using spectroscopic and photometric redshifts for the tracer galaxies. Then, in Sec.~4 we apply the algorithm to DES-SV data and discuss the choice of redshift slices and the way we deal with survey edge effects. Finally, in Sec.~5 we use the final DES-SV void catalog to measure the weak gravitational lensing around voids and we discuss our results and conclusions in Sec.~6.

\section{Data and simulations}
\label{sec:data}

The Dark Energy Survey (DES, \citealt{Flaugher2005,Flaugher2015,DarkEnergySurveyCollaboration2016}) is a photometric redshift survey that will cover about one eighth of the sky (5000 sq. deg.) to a depth of $i_{AB} < 24$, imaging about 300 million galaxies in 5 broadband filters ($grizY$) up to redshift $z=1.4$. The DES camera (DECam, \citealt{Flaugher2015}) includes sixty-two 2048x4096 science CCDs, four 2048x2048 guider CCDs, and eight 2048x2048 focus and alignment chips, for a total of 570 megapixels. In this paper we use 139 sq.~deg.~of data from the Science Verification (SV) period of observations \citep{Diehl2014}, which provided science-quality data at close to the nominal depth of the survey. 

In a photometric redshift survey, such as DES, the \pzs of tracer galaxies will impact the identification of voids with sizes comparable to the \pz scatter $\sigma_z$, in a way that renders some voids smeared and undetected. For DES main galaxies, this is a problem since $\sigma_z \simeq 0.1$ \citep{Sanchez2014}, corresponding to $\sim 220$ Mpc/$h$ at $z=0.6$, and typical voids have a comoving size of about 10-100 \mpch. However, we do not need to use all DES galaxies as void tracers. Instead, we can restrict ourselves to the Luminous Red Galaxies (LRGs) in the sample, which are still good tracers of the large-scale structure and have much better \pz resolution. 

\subsection{Void tracer galaxies: the \redmagic~catalog}

The DES-SV \redmagic~catalog \citep{Rozo2015} presents excellent \photoz \, performance: \redmagic\ photometric redshifts are nearly unbiased, with median bias $(z_{\mathrm{spec}}-z_{\mathrm{phot}})\approx 0.5\%$, a scatter 
$\sigma_z/(1+z)\approx 1.70\%$, and a $\approx 1.4\%$ $5\sigma$ redshift outlier rate. That scatter corresponds to a redshift resolution of $\sim 50$ Mpc/$h$ at $z=0.6$, a substantial improvement over DES main galaxies. Next we summarize the \redmagic~selection algorithm, but we refer the reader to \citet{Rozo2015} for further details. 

The red-sequence Matched-filter Galaxy Catalog (\redmagic, \citealt{Rozo2015}) is a catalog of photometrically
selected luminous red galaxies (LRGs). We use the terms \redmagic\ galaxies and LRG interchangeably. Specifically, \redmagic\ 
uses the redMaPPer-calibrated model for the color of red-sequence galaxies as a function of magnitude and redshift \citep{Rykoff2014}. 
This model is used to find the best fit photometric redshifts for all galaxies under the assumption that they are red-sequence members, and the $\chi^2$ goodness-of-fit of 
the model is then computed. For each redshift slice, all galaxies fainter than some minimum luminosity threshold $L_{\rm min}$ are 
rejected. In addition, \redmagic\ applies a cut $\chi^2 \leq \chi_{\rm max}^2$, where the cut $\chi_{\rm max}^2$ as a 
function of redshift is chosen to ensure that the resulting galaxy sample has a constant space density $\bar{n}$. In this work, we 
set $\bar{n}=10^{-3} h^3 \rm{Mpc}^{-3}$ with $\Lambda$CDM cosmological parameters $\Omega_{\Lambda}=0.7$, $h_{0}=100$, and \redmagic~galaxies are selected in the redshift range $0.2<z<0.8$. 
We expect the redMaGiC galaxy selection to be only marginally sensitive to the cosmological parameters assumed (see \citealt{Rozo2015} for details).
The luminosity cut is $L\geq L_*(z)/2$, where the value of $L_*(z)$ at $z=0.1$ is set to 
match the redMaPPer definition for SDSS \citep{Rykoff2014}, and the redshift evolution for $L_*(z)$ is that predicted using a simple passive evolution 
starburst model at $z=3$ \citep{Bruzual2003}. 

We use the \redmagic\ sample because of the exquisite photometric redshift performance 
of the \redmagic\ galaxy catalog. Also, because void properties depend on the tracer sample used, the constant comoving density of \redmagic~tracers helps in assuring the resulting voids have similar properties. For example, the dark matter profile \citep{Sutter2014} and void bias \citep{Chan2014,Clampitt2016,Pollina2016} have been shown to depend on the tracer density or tracer bias used to define voids.

Aside from the data catalog presented above, in this work we also use $\Lambda$CDM simulations that mimic the properties of the DES-SV \redmagic~data set. The mock galaxy catalog is the Buzzard-v1.0 from the Blind Cosmology Challenge (BCC) simulation suite, produced for DES (Wechsler et al, in preparation).
These catalogs have previously been used for several DES studies (see e.g. \citealt{Chang2015, Leistedt2015, Becker2015,Clampitt2016a,Kwan2016}). The underlying N-body simulation is based on three cosmological boxes, a 1050 Mpc/$h$ box with $1400^3$ particles, a 2600 Mpc/$h$ box with $2048^3$ particles and a 4000 Mpc/$h$ box with $2048^3$ particles, which are combined along the line of sight producing a light cone reaching DES full depth.
These boxes were run with LGadget-2 \citep{Springel2005a} and used 2LPTic initial conditions \citep{Crocce2006} with linear power spectra generated with CAMB \citep{Lewis2002}.
ROCKSTAR \citep{Behroozi2013} was utilized to find halos in the N-body volumes.
The ADDGALS algorithm (\citealt{Wechsler2004},  \citealt{Busha2013}, Wechsler et al, in preparation) is used to populate the dark matter simulations with galaxies as a function of luminosity and color.  
ADDGALS uses the relationship between local dark matter density and galaxy luminosity, to populate galaxies directly onto particles in the low-resolution simulations.  This relationship is tuned to reproduce the galaxy--halo connection in a higher resolution tuning simulation, in which galaxies are assigned using 
subhalo abundance matching \citep[e.g.][]{Conroy2006, Reddick2013}, in this case matching galaxy luminosity to peak circular velocity. Finally, each galaxy is assigned a color by using the color-density relationship measured in the SDSS \citep{Aihara2011} and evolved to match higher redshift observations.
The \redmagic~algorithm has been run on the simulation in a similar way as it is run on the DES data.
This produces a simulated sample with the same galaxy selection and photometric redshift performance as the DES-SV \redmagic~catalog but gives us access to the true redshifts of the galaxies in the sample, a fact that we will use to test the void finder presented in this work. 

\subsection{Lensing source catalog}

The catalog of galaxy shapes used in the lensing measurement of this work is the \texttt{ngmix}\footnote{https://github.com/esheldon/ngmix} catalog presented in \citet{Jarvis2015}. \texttt{ngmix} is a shear pipeline which produces model fitting shape measurements, and that was applied to a large subset of DES-SV galaxies, meeting the requirements of an extensive set of null and systematics tests in \citet{Jarvis2015}. The photometric redshifts of the galaxies in the \texttt{ngmix} shear catalog were studied in detail in \citet{Bonnett2015}, using 4 different photo-$z$ codes. In this work we use the SkyNet photo-$z$ method, which demonstrated excellent performance in that comparison.

\section{Photo-z void finder algorithm}
\label{sec:finder}

In this Section we present a new void finder designed specifically to work on photometric surveys. We explain the algorithm and test its performance on simulations, providing validation for the results shown later in the paper.

\subsection{Void finder algorithm}
\label{sec:algorithm}

The void finder works by projecting galaxies in redshift slices and finding underdensities in the 2D angular distribution of galaxies in the given slices. If the line-of-sight width of the projected slice is sufficiently large,  at least about twice the photo-$z$ resolution, then most galaxies will still be assigned to the correct slice. Since the finder works by projecting all galaxies within a given slice onto a 2D surface, the line-of-sight position within the slice does not affect the results.

The void finder of \citet{Clampitt2014} also begins by treating each slice in isolation, but has the disadvantage that voids are required to be completely empty of galaxies near the center. Thus, \pz scatter, which moves a single galaxy between slices, can inappropriately break up a single large void into several smaller voids, or even result in no void being detected at all. To overcome this problem, we smooth the 2D projected galaxy density field in each slice and then voids are found from minima of the smoothed density field. This means a few galaxies moving between different slices will not greatly affect the resulting set of voids, as will be demonstrated in Sec.~\ref{sec:sims}.

\begin{figure}
\centering
\includegraphics[height=80mm]{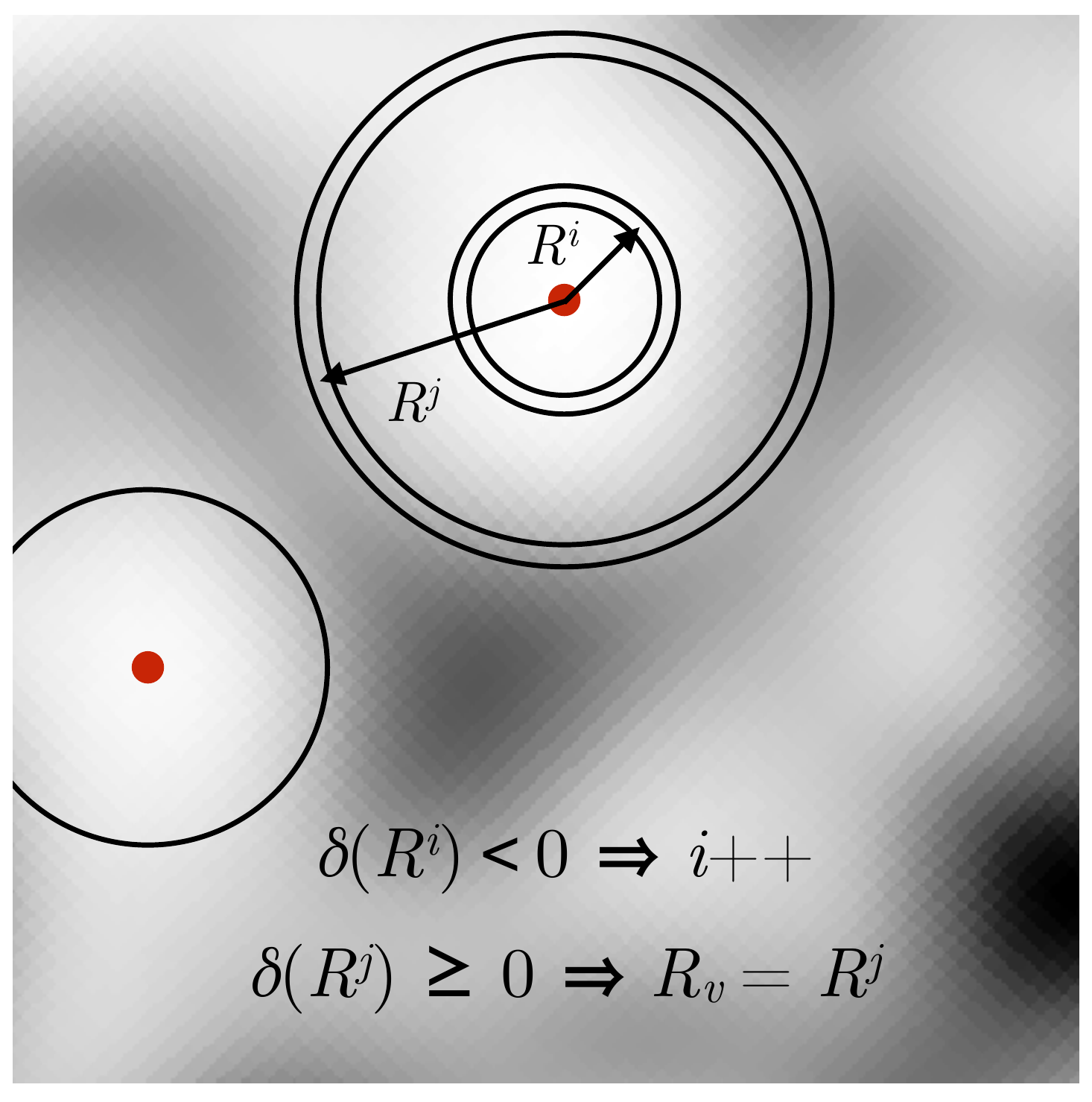}
\caption{Graphical description of the void-finding algorithm presented in this paper. The background gray-scaled field is the smoothed galaxy field ($\sigma = 10$ \mpch) in a redshift slice used by the void-finder. The two solid (red) dots show two void centers. For the upper void, we show a circular shell or radius $R^{i}$. Since the density contrast $\delta(R^{i}) < 0$, the algorithm checks larger shells, up to radius $R^{j}$ such that $\delta(R^{j}) \geq 0$. The void radius is then defined as $R_v = R^{j}$.}
\label{fig:finder}
\end{figure}

In detail, the void finding algorithm involves the following steps:

\begin{enumerate}[leftmargin=.2in]
\item We select the galaxies from a redshift slice of thickness $2s_v$ \cs{(we define $s_v$ to be half the slice thickness)} and we project them into a HEALpix map \citep{Gorski2005}, with a resolution of $N_{\textrm{side}} = 512$ representing an angular resolution of 0.1 deg.~and a physical resolution of 1.5 \mpch~at $z=0.3$ (3 \mpch~at $z=0.6$). 
\item We compute the mean density in the map corresponding to the given redshift slice, $\bar{n}_{2d}$, and convert the galaxy map to a density contrast map as $\delta = n_{2d}/\bar{n}_{2d} - 1$, where $n_{2d}$ is the galaxy map. 
\item Then we smooth the density contrast map with a Gaussian filter of comoving scale $\sigma_s = 10$ \mpch.
\item We take this smoothed contrast map and consider only the most underdense pixels (with $\delta < \delta_m = -0.3$) as potential void centers. We define the most underdense pixel in the map as the first void center.
\item Next we start defining circular shells of increasing radius around that center, stopping when the mean density within the slice ($\delta = 0$) is reached. That is, starting with a shell of radius $R_v^{\: i}$, we measure the average galaxy density in the shell $\delta(R_v^{\: i})$, and if the density is negative we check the next larger shell $\delta(R_v^{\: i+1})$, where the increment between shells is 1 Mpc/$h$ in radius. For some shell $R_v^{\: j}$ the density contrast reaches zero, $\delta(R_v^{\: j}) \geq 0$, and at that point the void radius is defined as $R_v = R_v^{\: j}$ (see Fig. \ref{fig:finder} for a graphical explanation).
\item Then all pixels contained in this void are removed from the list of potential void centers, \cs{preventing any of these pixels to become the center of any other void}. From the remaining pixels that satisfy $\delta < \delta_m = -0.3$, we define the next most underdense pixel as the second void center. The process is repeated until all pixels with $\delta < \delta_m = -0.3$ have been assigned to a void.
\end{enumerate}

Beyond the dependency on the line-of-sight size of the projected slice in which the finder is executed, studied in more detail later in this section, the void catalog produced by this algorithm depends on two parameters: the smoothing scale, $\sigma_s$, and the maximum density contrast of a pixel to become a void center, $\delta_m$. The smoothing scale ($\sigma_s = 10$ \mpch) is chosen to be about half the radius of the smallest voids we can access in our data sample (because of \pz smearing), and increasing it would erase the structure leading to some of these smallest voids, leaving the large voids intact. \textcolor{black}{On the other hand, the most significant voids found by the algorithm, the deepest ones, are independent of the choice $\delta_m = -0.3$ since their void center pixel is more underdense than that. By changing the value of $\delta_m$ we are only affecting the shallower voids of the sample. The impact of the $\delta_m$ choice is studied in Appendix \ref{sec:appendix_new}.} Also, voids found by this algorithm can overlap or even enclose one another, but just in the case where a subvoid is deeper than the bigger void enclosing it.

The process detailed above will produce a list of voids for a given redshift slice.
Before describing how various slices are combined to obtain the full void catalog, we first study the performance of the single slice results in simulations.

\subsection{Performance on simulations} 
\label{sec:sims}

\begin{figure*}
\centering
\includegraphics[width=180mm]{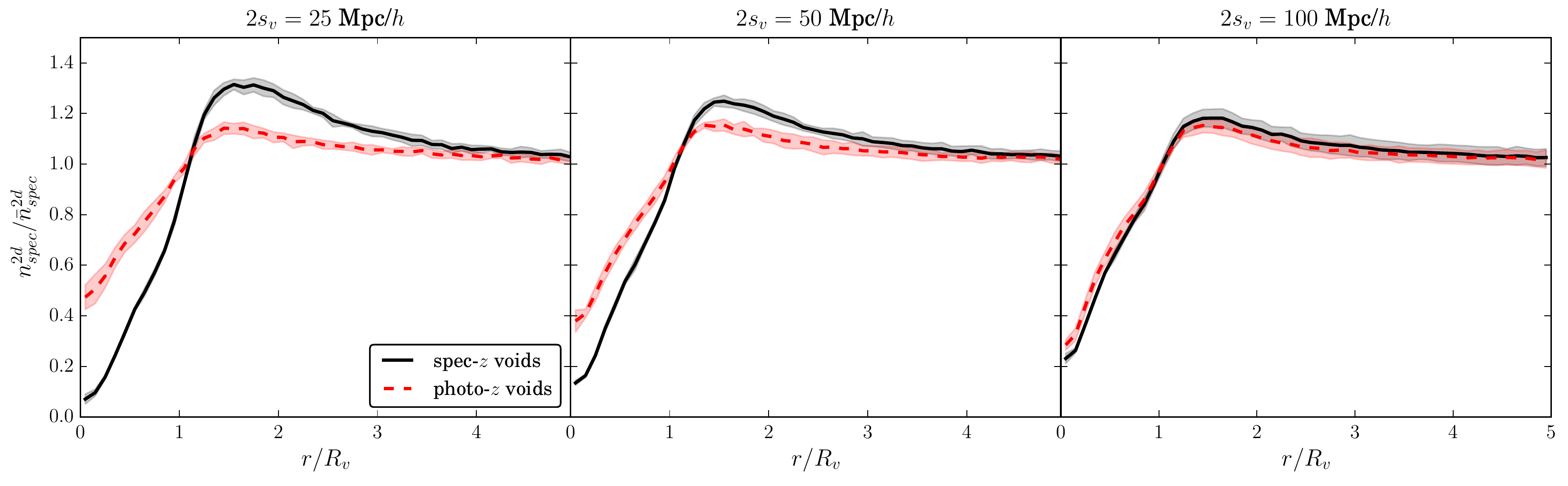}
\caption{({\it left panel}): Comparison of 2D spectroscopic galaxy density profiles of voids found in the simulations using galaxy spectroscopic redshifts (solid line) or photometric redshifts (dotted, red). \cs{The shaded regions show the corresponding error bars computed as the standard deviation among all the stacked voids.} The projected 2D slice width is 25 \mpch (comoving distance), a scale corresponding to $\sim 1/2$ the photometric redshift scatter. For this thin slice, the galaxy density profile is damped significantly by photo-z scatter, making the galaxy profile of \pz defined voids more shallow.
({\it center panel}): The same, but for a thicker slice of width 50 \mpch, comparable to the \pz scatter.
({\it right panel}): The same, but for a projected slice of width 100 \mpch, twice the size of the typical \pz scatter. In this case there is a good match between the profiles of spec-$z$ and \pz selected voids.
For such a thick slice, the fraction of galaxies that are placed in the incorrect slice due to photometric redshift scatter is smaller, allowing accurate void identification from the smoothed galaxy field.} 
\label{fig:profiles}
\end{figure*}

In order to validate the performance of the algorithm we use the simulations, where we have both spectroscopic and photometric redshift for void tracer galaxies, and we compare the voids found by the algorithm in spec-$z$ and photo-$z$ space. In particular, we run the void finding algorithm twice on each redshift slice: first using spectroscopic redshifts for selecting the galaxies that go into the slice and then using photometric redshifts that mimic the ones we have in real DES data.

Once we have the spec-$z$ and photo-$z$ defined void catalogs, we measure the projected galaxy density profiles of the voids in them in radial annuli using the true redshifts. Figure \ref{fig:profiles} shows the resulting density profiles for both cases in different slice comoving thicknesses. As expected, the void finder performs poorly if the size of the projected slice is smaller or similar to the \pz dispersion $\sigma_z \simeq 50$ \mpch. Therefore, the accuracy of the finder is a function of the thickness of the projected slice: for slice width $\sim 2$ times the size of the typical photometric redshift scatter, the difference between the average density profiles of voids found in spec-$z$ and \pz is not significant, being smaller than the standard deviation of the stacked void profiles.

\begin{figure}
\centering
\includegraphics[height=80mm]{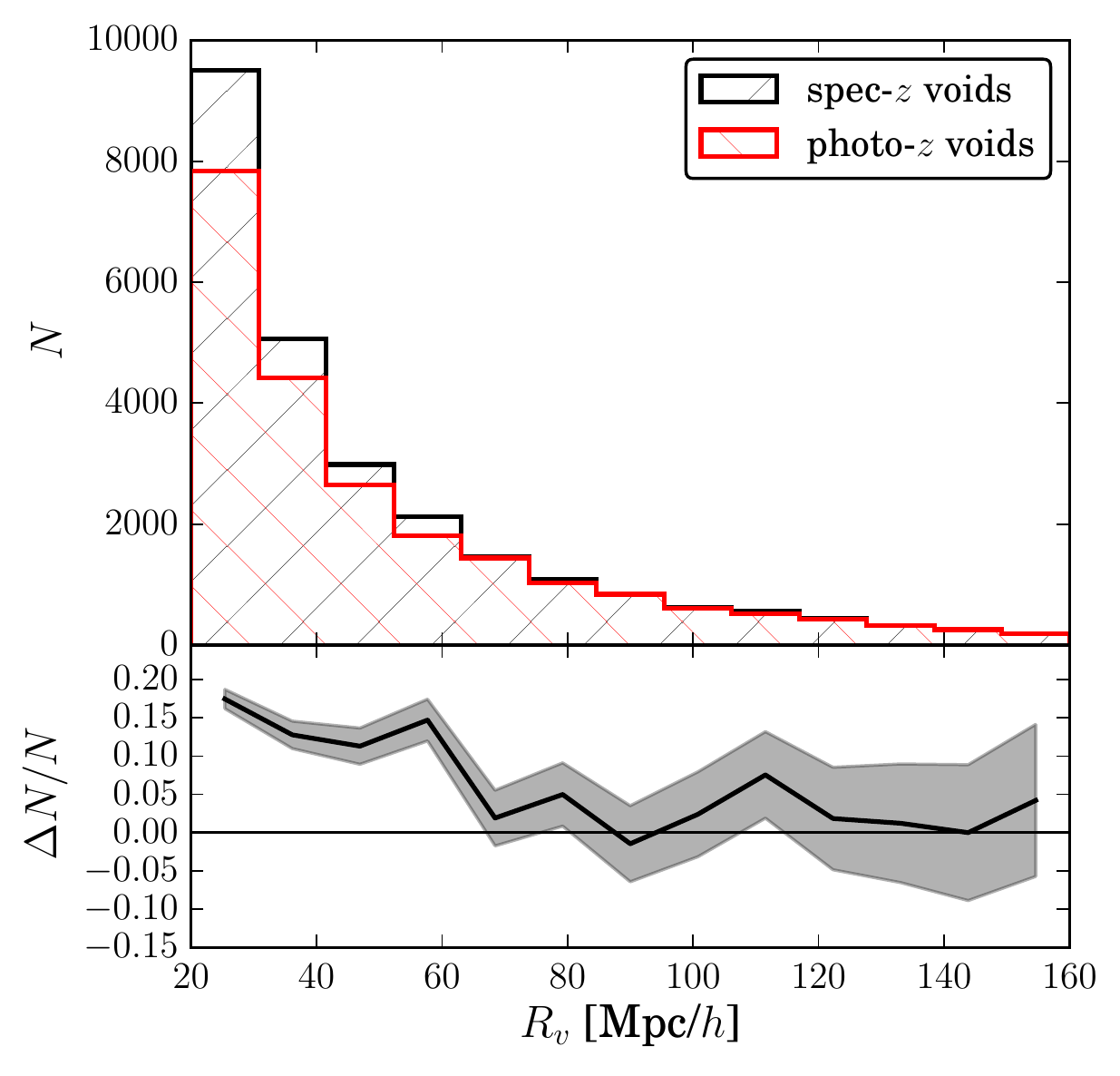}
\caption{({\it upper panel}): Void radius distribution for voids found in spec-$z$ and {\pz} simulated galaxy samples, for a slice thickness of $2s_v = 100$ \mpch.
({\it lower panel}): Relative difference between the distributions (with respect to the spectroscopic redshift case). Some voids with size smaller than the \pz scatter ($\sigma_z \simeq 50$ \mpch) are smeared out due to \pz scatter and not detected, resulting in a smaller number of voids relative to the spectroscopic case. For large voids this effect is not important and the two distributions agree within errors.}
\label{fig:rv_hist_sims}
\end{figure}

Figure \ref{fig:profiles} shows that voids found by the algorithm in \pz space can indeed have very similar density profiles as voids found in spec-$z$ space. However, it is also important to know the relative number of voids found in the two cases. Photometric redshifts produce a smearing in the line-of-sight position of tracers that can actually erase some of the structure, especially on scales comparable to the size of the \pz scatter or smaller. That will have the consequence of some small voids not being detected in the photo-$z$ case. The voids of size larger than the \pz scatter should be detected in both cases. Figure \ref{fig:rv_hist_sims} shows the distribution of void radii in simulations for spec-$z$ and photo-$z$ samples. As expected, we find less voids in the \pz case, with the difference being more important for small voids and becoming negligible for the voids substantially larger than the \pz dispersion ($\sigma_z \simeq 50$ \mpch).

\begin{figure}
\centering
\includegraphics[height=90mm]{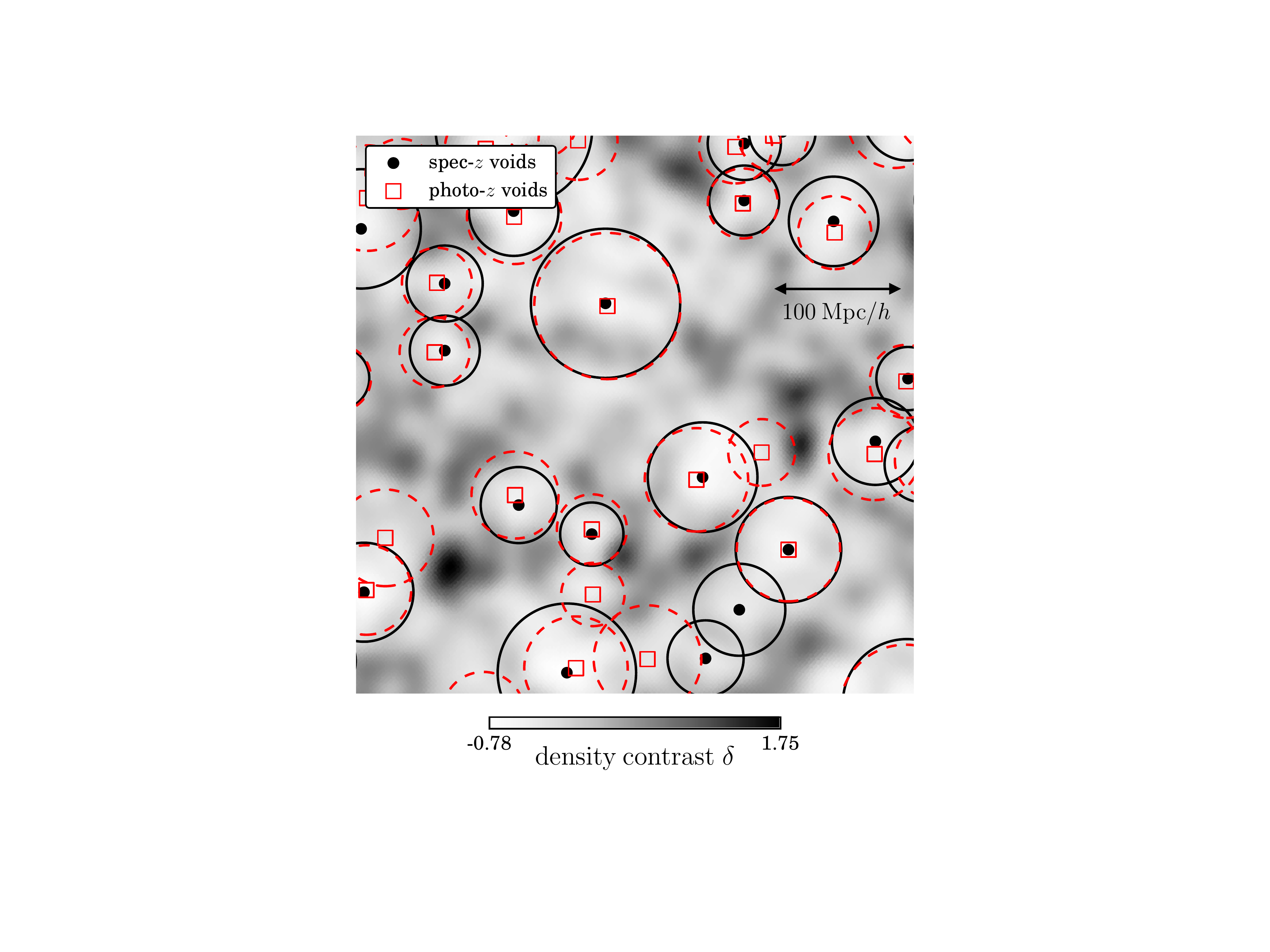}
\caption{Comparison between voids found in spec-$z$ (centers: solid black points; radius: solid circles) and \pz (centers: open red squares; radius: red dashed circles) in the simulations for a slice of thickness $2s_v = 100$~\mpch. The background gray-scaled field is the smoothed galaxy field ($\sigma = 10$ \mpch) used by the void-finder. The correlation between spec-$z$ and \pz defined voids is clear: in many cases the void position and radius match almost exactly.}
\label{fig:comparison}
\end{figure}

In addition to the comparison of the galaxy density profiles of voids, which is the most important test of the algorithm, Fig.~\ref{fig:comparison} shows a visual comparison between the positions and radius of spec-$z$ and \pz defined voids in a random $100$ \mpch-thick slice of our simulations. The correlation between the two sets of voids is very clear, in both positions and radii. In some cases, especially for the biggest voids, the match between spec-$z$ and photo-$z$ voids is almost perfect.
This is remarkable given the magnitude of the scatter in the line-of-sight direction being added by photometric redshifts.  

\section{DES-SV Void Catalog}
\label{sec:svcat}

In the previous Section we have presented a void finder that works by projecting galaxies into redshift slices \cs{(see Sec.~\ref{sec:algorithm} for a detailed description and parameters used in the algorithm)}. We have shown (Sec.~\ref{sec:sims}) that as long as the thickness of the projected slice is large enough compared to the \pz scatter, using photometric redshifts for the position of void tracers works nearly as well as using spectroscopic redshifts. Nevertheless, the algorithm will find some voids that are not likely to correspond to voids in the dark matter density field. Such false voids may be due to a number of effects: (i) at the survey edge or masked areas we have no information on galaxy positions, and (ii) duplicate voids may appear if slices overlap in redshift. In this Section we apply the algorithm to real DES-SV data, and present the way we deal with voids near the survey edge (Sec.~\ref{sec:pruning}) and the strategy we follow to get the most of the line-of-sight information in the data (Sec.~\ref{sec:kmeans}). \cs{The properties of the final DES-SV void catalog are presented in Sec.~\ref{sec:final_cat}.}

\subsection{Voids near the survey edge}
\label{sec:pruning}

\begin{figure}
\centering
\includegraphics[height=80mm]{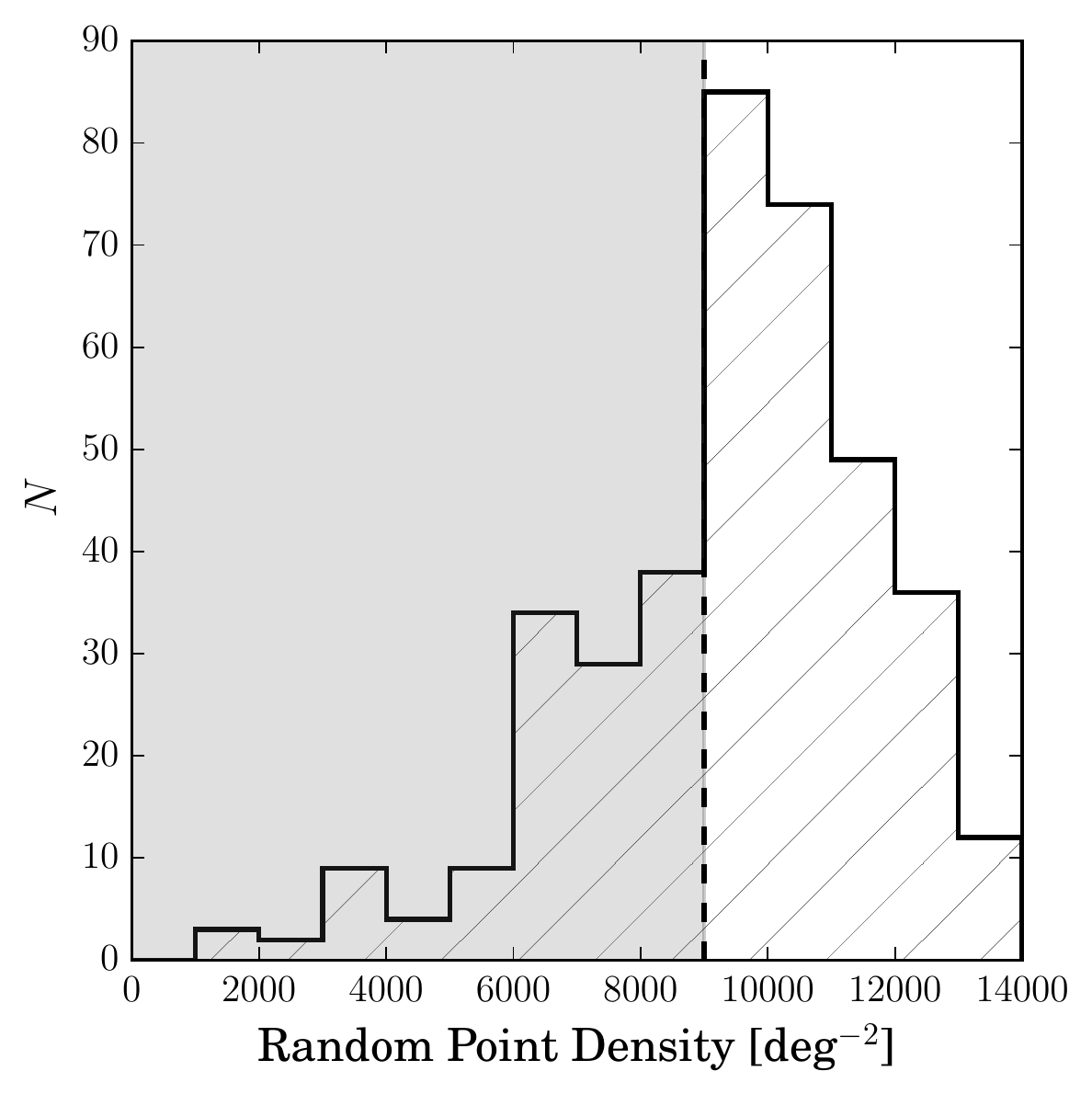}
\caption{Distribution of random point density inside DES-SV voids, where the random points are distributed uniformly through the DES-SV area. The distribution shows roughly a Gaussian shape at high densities corresponding to voids inside the survey mask, and a low density tail corresponding to edge voids. We remove all voids with random point density less than 9000 points/deg$^2$ (shaded region), most of them near the survey edge. This cut removes 33\% of the total number of voids.}
\label{fig:rp_hist}
\end{figure}

The assignment of each void's radius does not distinguish between voids that are fully contained within the survey and those that extend beyond it. The void radius may stretch beyond the edge of the survey, into areas which may or may not correspond to voids in the galaxy distribution. To remove such voids which extend far beyond the survey edge, we use the method of \citet{Clampitt2014}. A random point catalog drawn using the survey mask is generated, and for each void we calculate the density of random points inside $R_v$. The distribution of random points density inside voids is shown in Fig.~\ref{fig:rp_hist}, and it presents a Gaussian-like shape at high densities (peaked around 9500 points/deg$^2$ with $\sigma \simeq$ 2000 points/deg$^2$), corresponding to voids centered in the survey mask, and a low density tail reaching almost zero density, which corresponds to edge voids. Due to the small size of the DES-SV patch used in this work, with an area of 139 sq.~deg., and the size of some of the voids detected (a void with $R_v \sim 80$ Mpc/$h$ would span more than 10 deg.~in diameter at $z=0.3$), we place a conservative cut and discard voids with random point density less than 9000 points/deg$^2$, which constitute 33\% of the total number of voids.

\begin{figure}
\centering
\includegraphics[height=80mm]{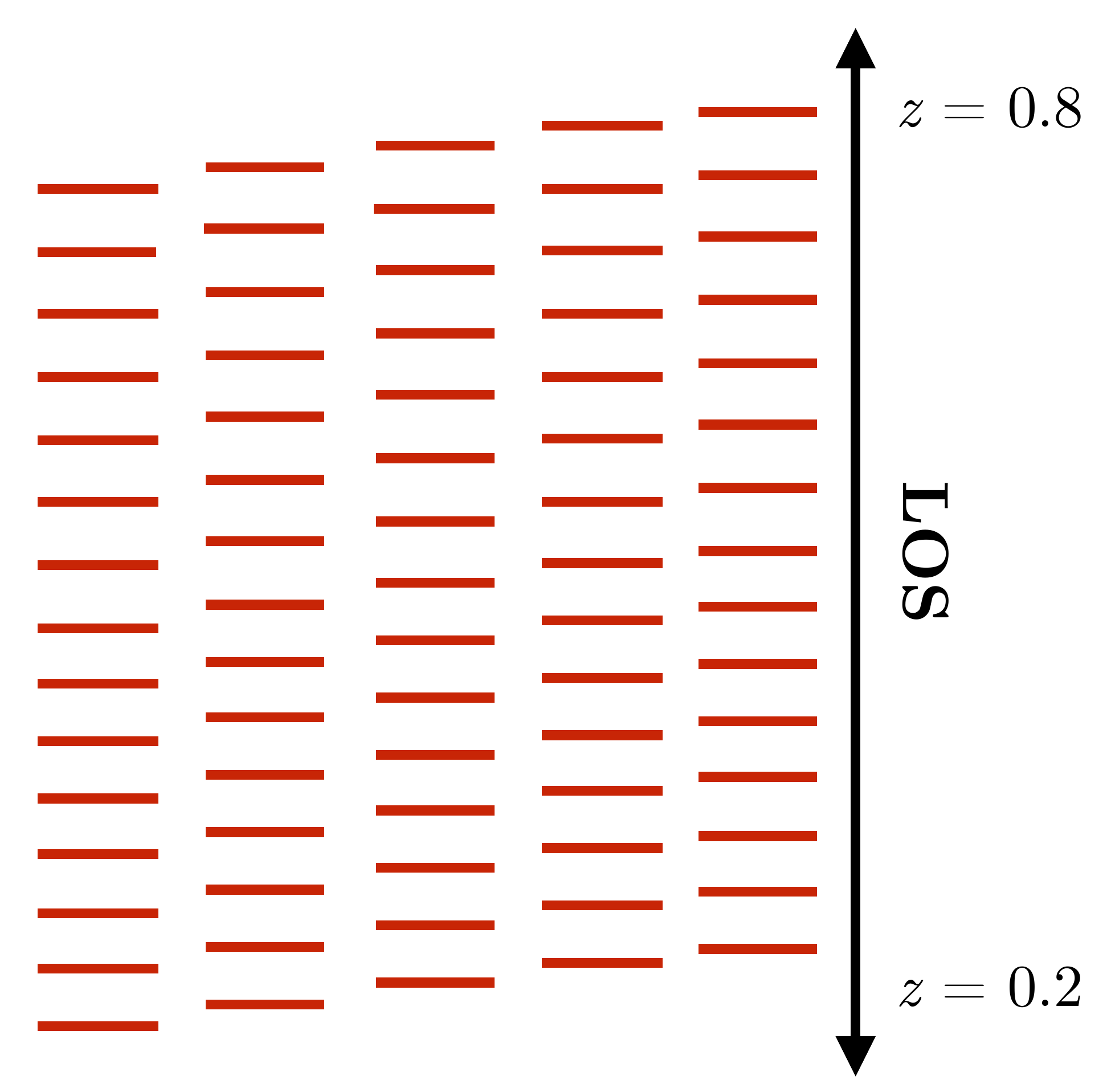}
\caption{Graphical representation of the line-of-sight (LOS) slicing performed in this paper. The black vertical arrow represents the redshift range, $0.2 < z < 0.8$, and the red horizontal bars represent the boundaries of the redshift slices in which the void finder is run. As the diagram shows, we oversample the line of sight with slices of thickness 100 \mpch~every 20 \mpch.
In Fig.~\ref{fig:kmeans} we show the way voids in adjacent slices are combined to form the final catalog.}
\label{fig:los}
\end{figure}

\begin{figure*}
\centering
\includegraphics[width=180mm]{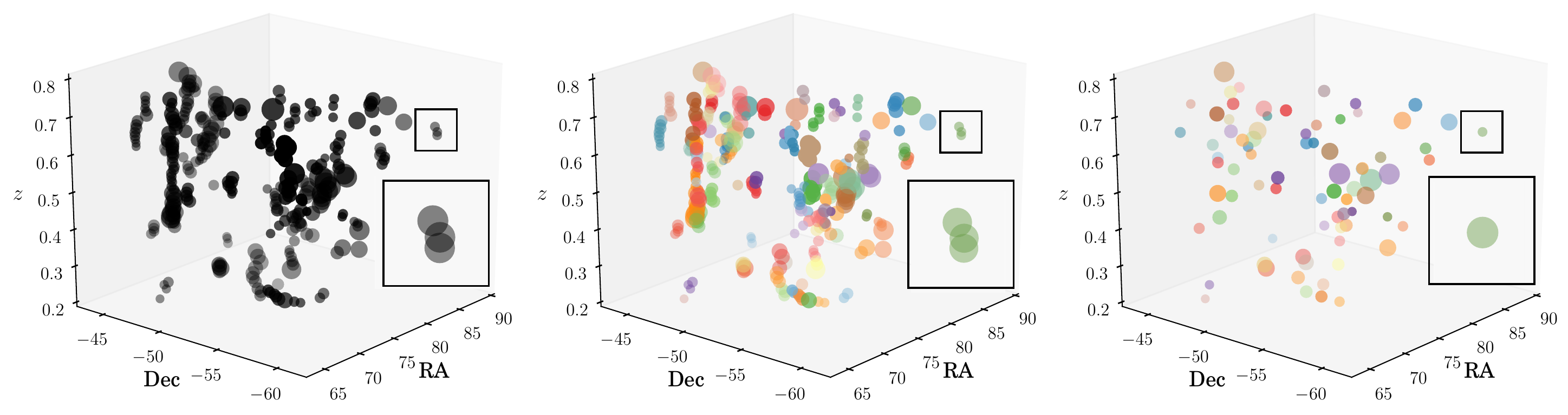}
\caption{(\textit{left panel}): 3D position of voids found in the slicing shown in Fig.~\ref{fig:los}. \cs{Each void candidate is shown as a sphere with size proportional to the void radius}. Due to oversampling in the line of sight, slices overlap and duplicates of the same physical void are found in different slices, apparent in this plot as elongated structures in redshift. The inset square shows the case of a three void group. (\textit{center panel}): Voids corresponding to the same physical underdensity are grouped together (as described in Sect. \ref{sec:kmeans}) and plotted with a common color. (\textit{right panel}): The final void positions are computed as the median 3D position of the members of each group.}
\label{fig:kmeans}
\end{figure*}

\subsection{Line of sight slicing strategy} 
\label{sec:kmeans}

To obtain more information about the line-of-sight position of each void we oversample the volume with a number of different slice centers. In particular, first we slice the line-of-sight range of the survey, $0.2 < z < 0.8$, in equal slices of comoving thickness $2s_v = 100$ \mpch~taking the upper redshift limit, $z=0.8$, as the upper limit of the furthest slice. Then we apply a shift to this slicing of 20 \mpch~towards low redshift, and we repeat the process four times so that we have a slice of thickness 100 \mpch~centered every 20 \mpch~of the line-of-sight range in the data (see Fig.~\ref{fig:los} for a graphical representation). 

\begin{figure}
\centering
\includegraphics[height=80mm]{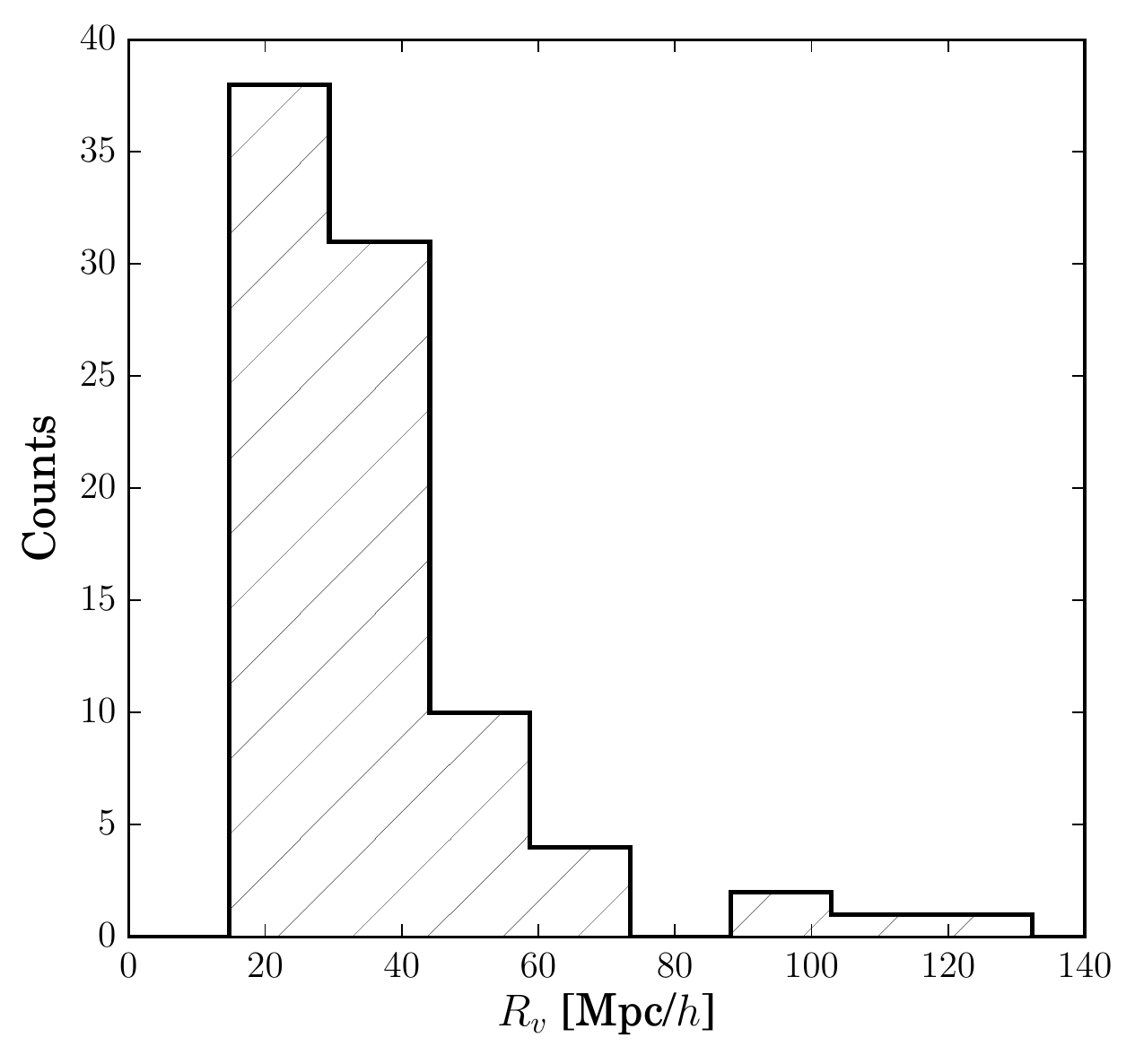}
\caption{\cs{Distribution of comoving void radii of the final DES-SV void catalog used in this work, using slices of thickness 100 Mpc/$h$ and after the cuts described in Sections \ref{sec:pruning} and \ref{sec:kmeans}.}}
\label{fig:rv_hist}
\end{figure}

\begin{figure}
\centering
\includegraphics[height=75mm]{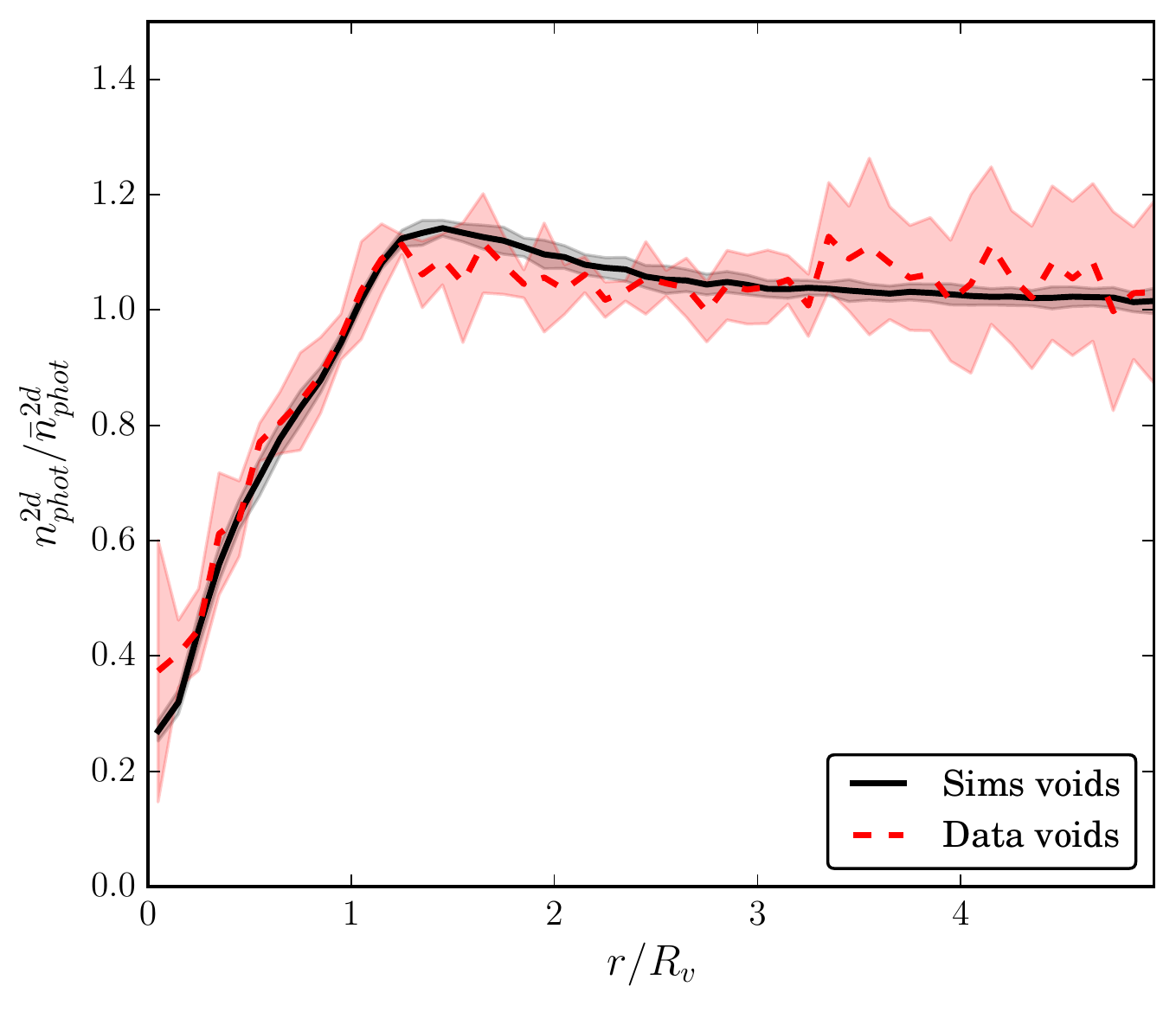}
\caption{\textcolor{black}{Comparison of 2D galaxy density profiles of voids found in DES-SV data and simulations, using galaxy photometric redshifts. The shaded regions show the corresponding error bars computed as the standard deviation among all the stacked voids.}}
\label{fig:profiles_data}
\end{figure}

Since the volume has been oversampled with a number of different slice centers, sometimes the same physical void will be found in multiple slices, creating elongated void structures in the line of sight (left panel in Fig.~\ref{fig:kmeans}). Each of these structures \cs{may} actually correspond to one physical underdensity\cs{, or at least their void candidate members will have a consistent lensing profile since they are essentially at the same redshift and have very similar sizes}. In order to remove the duplicate voids, and also to pick up the right void center in the line-of-sight direction, we need to group these void structures together. The groups are found by joining voids in neighboring (and hence overlapping) slices that have a small angular separation between them. In particular, two voids with radii $R_v^i$ and $R_v^j$ and found in neighboring slices will become part of the same group if the angular distance between their centers is smaller than half the mean angular radii of the two voids: $\bar{R_v}/2 = (R_v^i + R_v^j)/4$. The groups are shown in the central panel in Fig.~\ref{fig:kmeans}\cs{, and the right panel shows the final void catalog, without obvious elongated structures in the line of sight. This resulting void catalog is not very sensitive to the choice of $\bar{R_v}/2$: Increasing this minimum separation from $0.5 \bar{R_v}$ to $0.6 \bar{R_v}$ ($0.8 \bar{R_v}$) results in removing 6\% (10\%) of the voids in the final catalog.}

Once we have the void groups corresponding to those line-of-sight structures, we compute the 3D position of each group (RA, Dec and redshift) as the median position of the different void members of the group. The relative scatter in this determination inside each group (taken as the standard deviation of each quantity with respect to its mean value) is very small (less than 0.4\% for RA and Dec and around 2\% in redshift). The void radius is also computed as the median void radius of the different void members in each group, with a relative scatter around 14\%. The final void candidates, after removal of duplications of potential physical underdensities due to the oversampled slicing, are shown in the right panel of Fig.~\ref{fig:kmeans}. The effect of the LOS slicing strategy in the void lensing measurement is tested in Appendix \ref{sec:appendix_b}, where we show it helps reduce the noise but it does not affect the main outcomes from the measurement. 

\subsection{Final void catalog} 
\label{sec:final_cat}

\cs{Applying the void finding algorithm described in Sect.~\ref{sec:finder}, using slices of 100 Mpc/$h$ thickness, to the DES-SV \redmagic~catalog, and after making the cuts presented in Sections \ref{sec:pruning} and \ref{sec:kmeans}, we find a total of 87 voids in the 139 sq.~deg.~of survey area. These voids are identified in the redshift range $0.2<z<0.8$, and they have comoving sizes ranging from $R_v = 18$ Mpc/$h$ to $R_v = 120$ Mpc/$h$, with a mean void radius of $\bar{R_v} = 37$ Mpc/$h$. Figure \ref{fig:rv_hist} shows the full void radius distribution for the sample. The mean angular radius of voids in the sky is 1.5 degrees while their mean redshift is $\bar{z} = 0.57$. }

\textcolor{black}{Figure \ref{fig:profiles_data} shows the 2D galaxy density profiles of voids found in the DES-SV data and in simulations, using galaxy photometric redshifts. The agreement between data and simulations is good, and so is the agreement between the simulation profiles measured with photometric (Fig.~\ref{fig:profiles_data}) and spectroscopic redshifts (\textit{right panel} of Fig.~\ref{fig:profiles}).}

\section{Void Lensing}
\label{sec:lensing}

Using the void catalog defined in the previous section we now focus on the lensing measurement around voids. This represents a key result, since a significant lensing signal around voids proves them to be underdense in the matter field, this way demonstrating the void catalog is primarily composed of real underdensities rather than spurious detections, tracer density effects or any systematics in the data.

In this section we present the details of the lensing measurement and covariance, the results for the tangential and cross components of that measurement and their significance, and the fit of the tangential component to a void model widely used in the literature. 

\subsection{Measurement}
\label{sec:measurement}
\begin{figure}
\centering
\includegraphics[height=80mm]{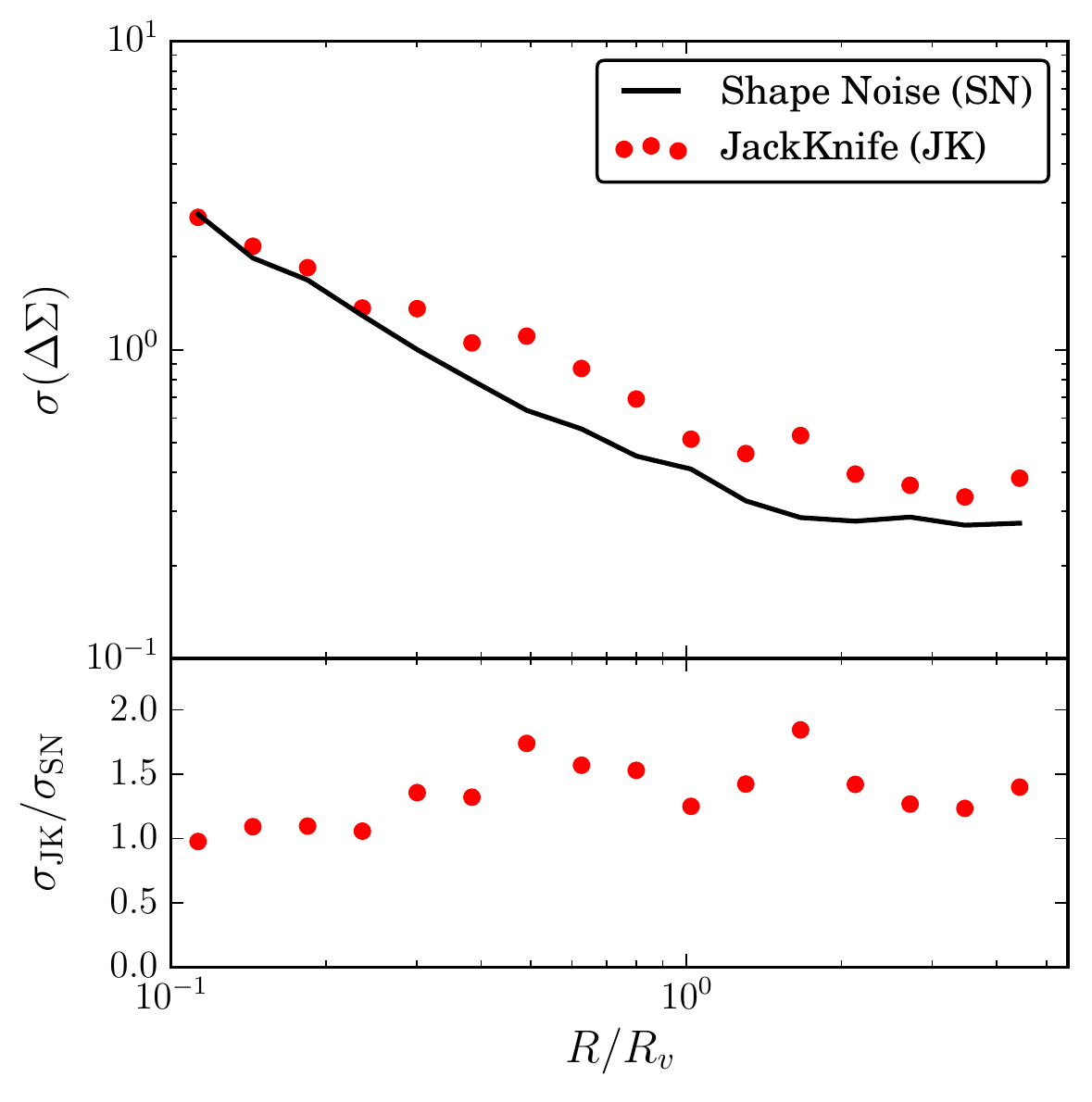}
\caption{\textit{(Upper panel):} Variance in the stacked weak lensing measurement of voids in DES-SV data, in bins of $R/R_v$, as estimated from jackknife (JK) resampling and lensing shape noise, the two techniques described in Sect.~\ref{sec:covs}. \textit{(Lower panel):} Ratio of the two error estimations in the upper panel. The two agree well on small scales (which are shape noise dominated) and differ significantly at medium to large scales since the jackknife includes other sources of variance in addition to shape noise. } 
\label{fig:cov_comparison}
\end{figure}

Assuming an axisymmetric density profile, \cs{the stacked excess surface mass density} $\Delta\Sigma$ is related to the tangential shear $\gamma_t$ of source galaxies by
\begin{equation}
\Delta \Sigma (R/R_v) = \Sigma_{\rm crit} \gamma_t (R/R_v) \, ,
\end{equation}
where the proportionality factor describing the lensing strength is
\be
\Sigma_{\rm crit} (\zl, \zs) = \frac{c^2}{4\pi G} \frac{D_A(\zs) (1+\zl)^{-2}}{D_A(\zl) D_A(\zl,\zs)} \, ,
\ee
with $\Sigma_{\rm crit}^{-1}(\zl,\zs)=0$ for $\zs<\zl$, \cs{where $\zl$ and $\zs$ are the lens and source galaxy redshifts, respectively}.
Note both the use of comoving units and that we need to assume a certain cosmology (flat $\Lambda$CDM with $\Omega_m = 0.3$) when calculating the angular diameter distances $D_A$ in $\Sigma_{\rm crit}$.
Our lensing projected surface density estimator is therefore given by
\begin{equation}
 \Delta\Sigma_{k}(R/R_v;\zl)=\frac{\sum_j 
\left[
w_j \gamma_{k,j}(R/R_v) \Sigma_{{\rm crit}, j}(\zl, \zs)
\right]
}{\sum_j w_j}
\label{eq:dsig}
\end{equation}
where $k$ denotes the two possible components of the shear (tangential and cross), 
the summation $\sum_j$ runs over all the source galaxies
in the radial bin $R/R_v$, around every void position, 
and the \cs{optimal} weight for the $j$-th galaxy is given by \cs{\citep{Sheldon2004}}:
\be \label{eq:weight}
w_j = \frac{[\Sigma_{{\rm crit}, j}^{-1}(\zl, \zs)]^{2}}{\sigma_{\rm shape}^2 + \sigma_{{\rm m},j}^2} \, .
\ee
Here $\sigma_{\rm shape}$ is the intrinsic shape noise for each source galaxy, and $\sigma_{{\rm m},j}$ is the shape measurement error.
In Sec.~\ref{sec:model} we relate the differential surface density $\Delta\Sigma$ to the 3D void profile $\rho_{\rm v}$.

Note that since the projected void radius $R_v$ ranges from 20 to more than 100 \mpch, we stack the measured shear profiles in units of the void radius, $R/R_v$.
Stacking the profiles in physical distance would smooth out the stacked void density profiles and hence some of the signal would be lost.  

\subsection{Covariance}
\label{sec:covs}

In order to estimate the covariance for the $\Delta\Sigma(R)$ measurements in this work we combine two different approaches: we rely on the jackknife (JK) method to estimate the signal variance while we estimate the off-diagonal shape of the covariance from the lensing shape noise of the measurement \cs{\citep{Melchior2014}}. The main reason for that combination is the limitation in the JK technique due to the small number of voids ($\sim 100$) in our catalog, yielding very noisy off-diagonal correlations. However, we can obtain smooth shape-noise-only covariances by applying any number of random rotations to the ellipticities of source galaxies. Next we explain the precise combination of the two approaches. 

Due to the small number of voids in the DES-SV catalog, we perform a void-by-void jackknife:
we carry out the measurement multiple times with each void omitted in turn to make as many jackknife realizations as voids we have in the sample ($N$). Then, the variance of the measurement \citep{Norberg2009} is given by

\begin{equation} \label{eq:err_jk}
\sigma_{\mathrm{JK}}^2 (\Delta \Sigma_i) = \frac{(N-1)}{N} \times \sum\limits_{\mathrm{JK}-k=1}^N \left[(\Delta \Sigma_i)^{\mathrm{JK}-k} - \overline{\Delta\Sigma_i}\right]^2
\end{equation}
where the mean value is
\be \label{eq:avg}
\overline{\Delta\Sigma_i} = 
\frac{1}{N}
\sum\limits_{\mathrm{JK}-k=1}^N (\Delta\Sigma_i)^{\mathrm{JK}-k}\, ,
\ee
and 
$(\Delta\Sigma_i)^{\mathrm{JK}-k}$
denotes the measurement from the $k$-th JK realization and the $i$-th
spatial bin.

The shape noise (SN) covariance of the measurement is estimated by randomly rotating the orientation of each source galaxy ellipticity many times ($N_\mathrm{SN}=300$ in this analysis) and repeating the $\Delta\Sigma$ lensing measurement each time. Then the covariance is estimated as:

\begin{flalign} \label{eq:cov}
\mathrm{Cov}_{\mathrm{SN}} [\Delta \Sigma_{i}, & \Delta\Sigma_{j}] = \frac{1}{N_\mathrm{SN}}
& 
\nonumber \\
& 
\times \sum\limits_{\mathrm{SN}-k=1}^{N_\mathrm{SN}} \left[(\Delta \Sigma_i)^{\mathrm{SN}-k} - \overline{\Delta\Sigma_i}\right]
\left[(\Delta \Sigma_j)^{\mathrm{SN}-k} - \overline{\Delta\Sigma_j}\right]
\end{flalign}
where the mean value is
\be \label{eq:avg}
\overline{\Delta\Sigma_i} = 
\frac{1}{N}
\sum\limits_{\mathrm{SN}-k=1}^N (\Delta\Sigma_i)^{\mathrm{SN}-k}\, ,
\ee
and 
$(\Delta\Sigma_i)^{\mathrm{SN}-k}$
denotes the measurement from the $k$-th shape noise (SN) realization and the $i$-th
spatial bin.

Figure \ref{fig:cov_comparison} shows a comparison of the measurement variance estimated from jackknife and shape noise, following the techniques described above. The errors coming from the two approaches agree well on the smallest scales, as expected since the small-scale regime is dominated by shape noise. However, at mid to large scales ($R \sim 0.28 R_v$ and above) the JK errors get bigger than SN only, as they can trace other effects such as systematics in the data or sample variance. The shape noise calculation is, on the other hand, more adequate for off-diagonal elements of the covariance since it avoids the intrinsic noise limitation of the JK technique. Hence, in order to have a smooth covariance matrix with variance accurately estimated from JK, we follow the approach of fixing the shape of the covariance as given by the shape noise calculation, and renormalize it to the JK estimates of the variance:

\begin{equation}
\mathrm{Cov} [\Delta \Sigma_{i}, \Delta\Sigma_{j}] = \mathrm{Corr}_{\mathrm{SN}} [\Delta \Sigma_{i}, \Delta\Sigma_{j}] \sigma_{\mathrm{JK}} (\Delta \Sigma_i) \sigma_{\mathrm{JK}} (\Delta \Sigma_j)
\end{equation}
where $\mathrm{Corr}_{\mathrm{SN}} [\Delta \Sigma_{i}, \Delta\Sigma_{j}]$ is the shape noise correlation matrix (or reduced covariance) given by:

\begin{equation}
\mathrm{Corr}_{\mathrm{SN}} [\Delta \Sigma_{i}, \Delta\Sigma_{j}] = \dfrac{\mathrm{Cov}_{\mathrm{SN}} [\Delta \Sigma_{i}, \Delta\Sigma_{j}]}{\sigma_{\mathrm{SN}} (\Delta \Sigma_i) \sigma_{\mathrm{SN}} (\Delta \Sigma_j)}
\end{equation}

The approach of renormalizing a smooth covariance to a JK estimated variance has been used before in the literature, for example by \citet{Crocce2016}. 

\subsection{Null tests: Cross-component and \textit{randomized} voids}
\label{sec:null}

\begin{figure}
\centering
\includegraphics[height=75mm]{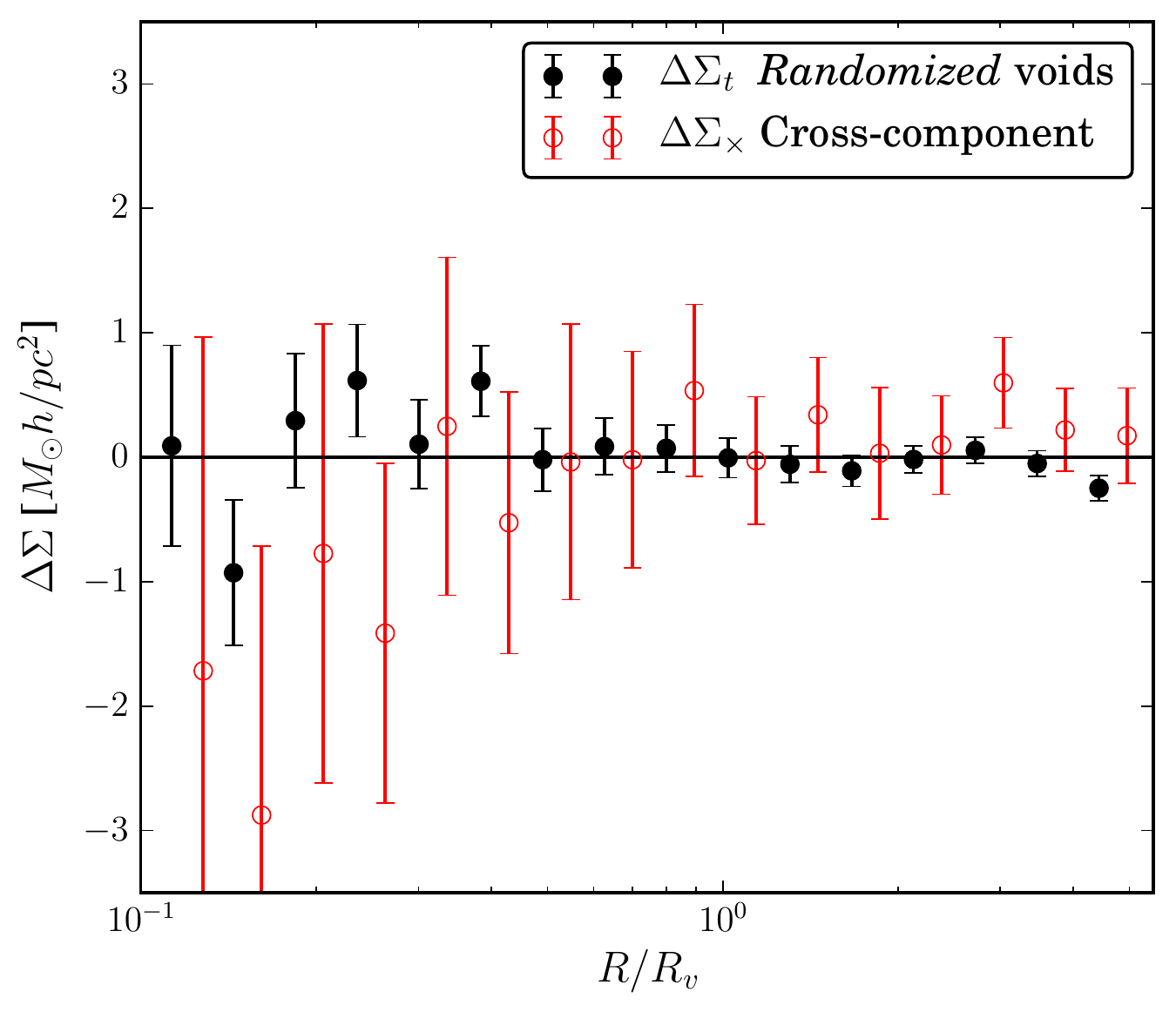}
\caption{Cross-component of the DES-SV data stacked lensing measurement for true voids and tangential component for the lensing around \textit{randomized} voids, in bins of $R/R_v$. Both measurements are compatible with the null hyposthesis with $\chi^2_{\mathrm{null}}/\mathrm{dof} = 8.2/16$ and $\chi^2_{\mathrm{null}}/\mathrm{dof} = 18.7/16$, respectively. The error using \textit{randomized} voids is smaller since the measurement involves $\sim$ 10 times more \textit{randomized} voids.}
\label{fig:null}
\end{figure}

The cross-component of the measurement described in Sect.~\ref{sec:measurement} is not produced by gravitational lensing and therefore is expected to 
vanish at first order. Similarly, the tangential component of the same measurement around \textit{randomized} voids, which follow the size and redshift distribution of true voids but are randomly distributed in the survey area (Appendix \ref{sec:appendix_a}), is also expected to vanish. Figure \ref{fig:null} shows the cross-component of the stacked lensing measurement for true voids and the tangential component for \textit{randomized} voids. 

With dof $= N_{\mathrm{bin}}$ as the number of $R/R_v$ bins in the measurement and no model parameters, the null hypothesis $\chi^2$ can be computed as
\begin{equation}
\chi^2_{\mathrm{null}} = \sum_{i,j} \Delta \Sigma_i \mathrm{Cov}^{-1}_{ij} \Delta \Sigma_j
\end{equation}
where $i,j$ correspond to radial bins in $\Delta\Sigma$ and $\mathrm{Cov}$ is the covariance matrix. 

\begin{figure}
\centering
\includegraphics[height=75mm]{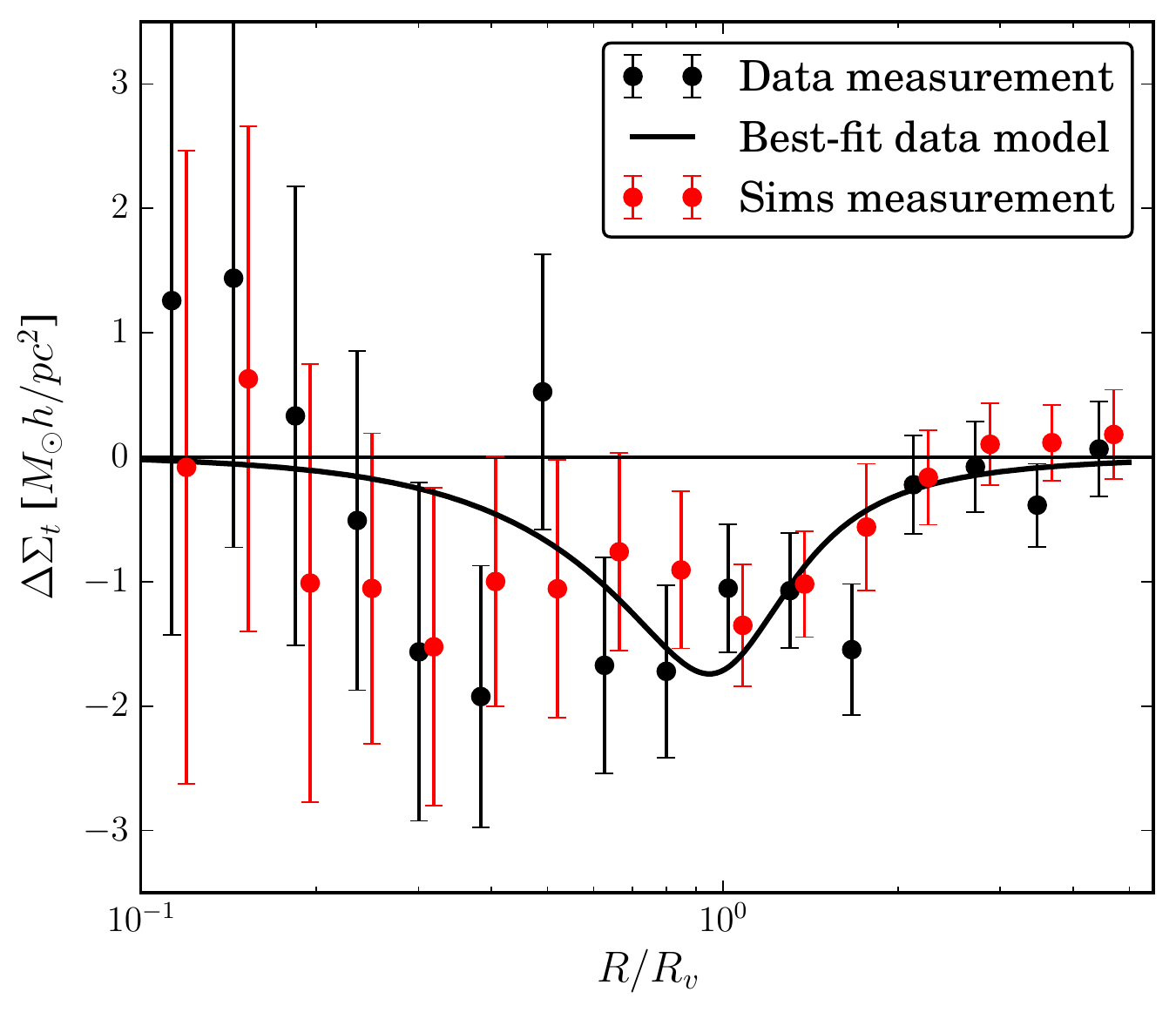}
\caption{\textcolor{black}{Stacked tangential shear profile around voids in DES-SV data (black points) and simulations (red points) in bins of $R/R_v$. The black solid line shows the best-fit model (see Sect.~\ref{sec:model}) to the data shear signal. The $\chi^2$ for the null hypothesis in the data measurement is $\chi^2_{\mathrm{null}}/\mathrm{dof} = 35.5/16$, yielding an estimated $S/N = 4.4$, while the theory model provides a good fit to the data with $\chi^2$/dof$ = 13.2/14$. The measurement in the simulations shows consistent with the data best-fit model, yielding $\chi^2$/dof$ = 10.1/14$.}} 
\label{fig:lensing_signal}
\end{figure}

The cross-component of the measurement yields a $\chi^2_{\mathrm{null}}/\mathrm{dof} = 8.2/16$, and the tangential measurement around \textit{randomized} voids, which are 10 times more numerous than true voids and whose production is described in greater detail in Appendix \ref{sec:appendix_a}, yields a $\chi^2_{\mathrm{null}}/\mathrm{dof} = 18.7/16$, both showing consistency with the null hypothesis. 

\subsection{Tangential shear profile}
\label{sec:results}

Figure \ref{fig:lensing_signal} shows the measurement of the tangential component of the stacked lensing signal around voids. Assuming a non-central $\chi^2$ distribution we can compute the signal-to-noise ($S/N$) of the measurement as
\begin{equation}
(S/N)^2 = \chi^2_{\textrm{null}} - \textrm{dof} = \sum_{i,j} \Delta \Sigma_i \mathrm{Cov}^{-1}_{ij} \Delta \Sigma_j - N_{\mathrm{bin}}
\label{eq:s2n}
\end{equation}
The evaluation of this expression yields $\chi^2$/dof = 35.5/16 and hence $S/N = 4.4$. The significance of the signal is complemented with the null tests in the previous subsection being consistent with the null hypothesis. Furthermore, we test the robustness of the signal to changes in the LOS slicing strategy in Appendix \ref{sec:appendix_b} and to changes in the value of $\delta_m$ in Appendix \ref{sec:appendix_new}. 

\subsection{Model fits}
\label{sec:model}

\begin{figure}
\centering
\includegraphics[height=75mm]{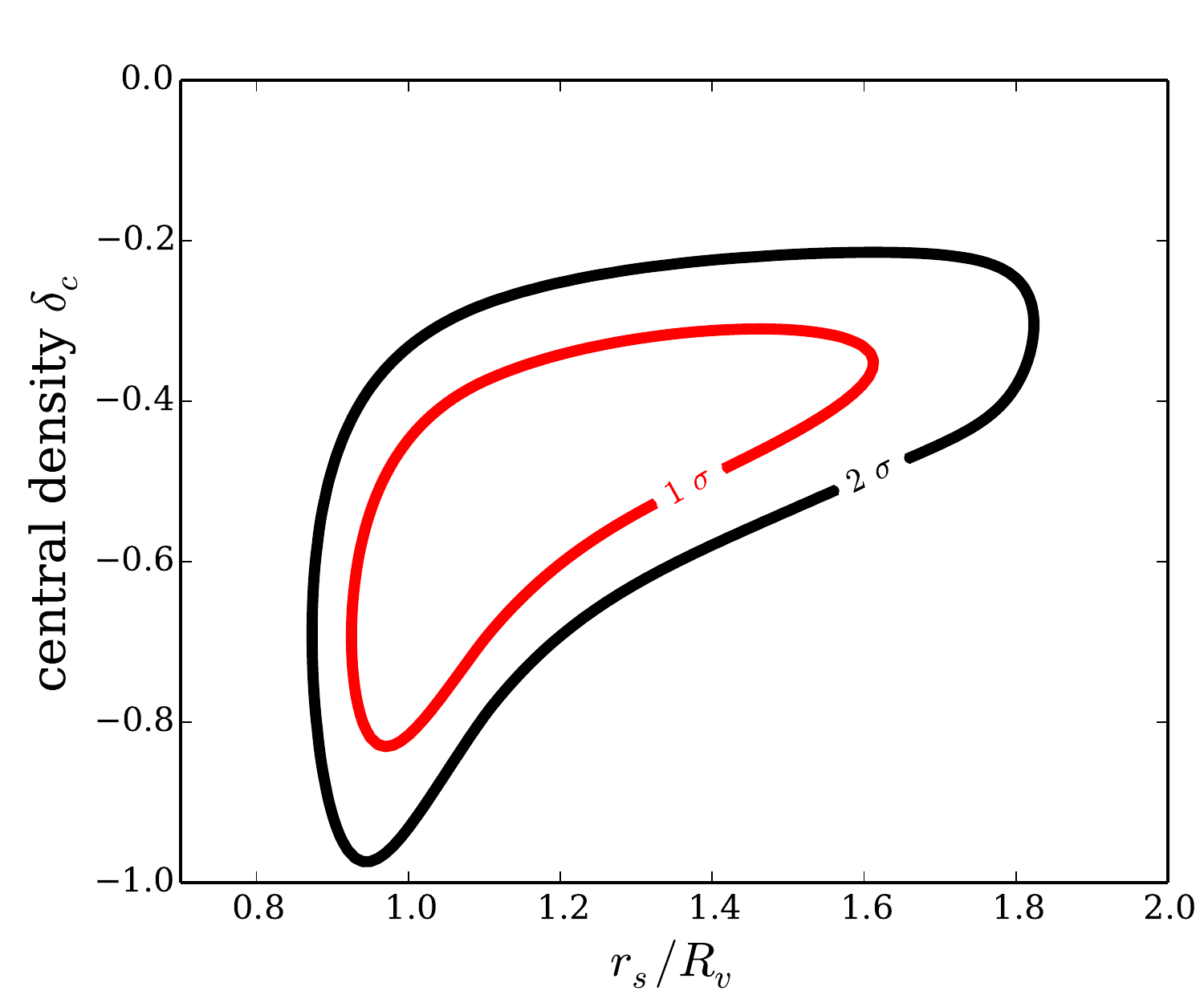}
\caption{Constraints on void central underdensity $\delta_c$ and scale radius $r_s$ from the DES-SV data void lensing measurements in Fig.~\ref{fig:lensing_signal}. Best-fit values are $r_s = 1.05 R_v$ and $\delta_c = -0.60$, and the $\chi^2$/dof for the fit is 13.2/14. There is good agreement between the void edge determined from galaxies, $R_v$, and the void edge determined from lensing, $r_s$.}
\label{fig:contours}
\end{figure}

We use the 3D void profile of \citet{Hamaus2014a} \cs{(henceforth HSW14)}
\be
\frac{\rho_{\rm v} (r)}{\bar{\rho}} -1 = \delta_c \frac{1 - (r/r_s)^\alpha}{1 + (r/R_{\rm v})^\beta} \, ,
\label{eq:model}
\ee
and fit two parameters: the central underdensity $\delta_c$ and the scale radius $r_s$.
Note that $r$ here denotes the 3D (in contrast to projected) radius.
We do not fit the inner and outer slopes $\alpha$ and $\beta$ using the lensing data, but fix their values to the simulation fits of HSW14.
That work showed that $\alpha$ and $\beta$ are not independent parameters but determined by the ratio $r_s / R_{\rm v}$, \cs{which yields $\alpha = 2.1$ and $\beta = 9.1$ for the best fit $r_s$ shown in Fig.~\ref{fig:contours}}.
Following Krause et al. (2013) the lensing observable $\Delta\Sigma(R/R_v)$ is related to the 3D density by
\be
\Delta\Sigma(R/R_v) = \bar{\Sigma}(<R/R_v) - \Sigma(R/R_v) \, ,
\ee
where the projected surface density is given by
\be
\Sigma(R/R_v) = \int {\rm d} r_{\rm los} \, \rho_{\rm v} \left(\sqrt{r_{\rm los}^2 + R^2}\right) - \bar{\rho} \, ,
\ee
and $\bar{\rho}$ is the cosmological mean mass density.

The resulting parameter constraints are shown in Fig.~\ref{fig:contours}.
The reduced $\chi^2/\textrm{dof} = 13.2 / 14$ implies a good fit to the theory model. 
\textcolor{black}{Even though the uncertainties are important, the best-fit $\delta_c = -0.60$ is in agreement with the density profile shown in Fig. 9, which is at the same time in agreement with the profile measured in simulations. In order to further support the data measurement using simulations, we have measured the lensing signal in the simulations using the same number of voids as in the data. The resulting measurement can be found in Fig~\ref{fig:lensing_signal}, and it shows consistency with the best-fit model to the data with $\chi^2/\textrm{dof} = 10.1 / 14$.} 

Additionally, the best-fit $\delta_c$ and the trend in Fig.~\ref{fig:contours} are in agreement with findings in HSW14. \textcolor{black}{However, note the important differences between our work and HSW14: we use photometric galaxies instead of N-body dark matter particles. More importantly, we are using a different void finder. Thus it should not be surprising that our mean void radius ($R_v$), scale radius ($r_s$), and mean void underdensity ($\delta_c$) do not match all the relations obeyed by theirs. For example, their void sample with $r_s/R_v \simeq 1.05$ (matching our best-fit value) is slightly smaller ($R_v \simeq 29$ \mpch) and more empty ($\delta_c \simeq -0.7$) than ours.}

Finally, we can use the constraints on $\delta_c$ being negative as an alternative estimate of the significance in the lensing detection, which is consistent with the estimation in eq.~(\ref{eq:s2n}): marginalizing over $r_s$, we find $\delta_c < 0$ with a significance of $4.6\sigma$ ($4.8\sigma$ if we fix $r_s$ to its best-fit value). The best-fit value of $r_s$ is compatible with $R_v$ at the 1-$\sigma$ level. Based on eq.~(\ref{eq:model}), $r = r_s$ is just the place where the local 3D density returns to the cosmic mean, $\rho = \bar{\rho}$. The definition of $R_v$ is based on where the local galaxy density returns to the mean (Fig. 1). So given this best-fit model we see that the void wall in the mass distribution (determined from lensing) agrees well with the void wall in the galaxy distribution.

\subsection{Comparison to previous measurements}

Other measurements of weak gravitational lensing around voids or underdensities have been performed in recent years. \citet{Melchior2014} used the SDSS void catalog of \citet{Sutter2012} to carry out the first detection of lensing around voids, although at low $S/N$. \citet{Clampitt2014}, using a similar data sample, optimized the void finding strategy for lensing purposes and were able to achieve a higher $S/N \sim 7$ in the lensing measurement. The void finder in this work is similar to that of \citet{Clampitt2014}, even though we did not attempt to optimise the lensing detection but to minimise the photo-$z$ related impact in the void finding procedure. Our comparable lensing $S/N$ is encouraging given the use of photometric redshifts and a smaller dataset -- this highlights the viability of photometric void finders as well as the quality of the DES data. 

\citet{Gruen2015} changed the approach and, instead of looking at individual cosmic voids, measured the lensing signal around troughs in the DES-SV galaxy distribution, defined as underdensities in the projection of lens galaxies over a wide range in redshift. That produced a high $S/N$ lensing measurement around those structures, and they succesfully modelled that to probe the connection between galaxies and matter. \cs{In that respect, trough lensing does not constrain void profiles or abundances but it is sensitive to the galaxy bias and even cosmology.}

\section{Discussion}
\label{sec:conclusions}

We have presented a new void finder designed for photometric surveys and applied it to early Dark Energy Survey data and simulations.
Fixing the line-of-sight size of the slice to be at least twice the photo-$z$ scatter, we find the number of voids found in simulated spectroscopic and photometric galaxy catalogs to be within 20\% for all transverse void sizes, and indistinguishable for voids with projected size larger than 70 \mpch.
For such large voids, most have a one-to-one match with nearly the same assigned center and radius.

This result -- that the largest voids are the ones most faithfully preserved in a photometric redshift survey -- has implications for the expected spatial and dynamic properties of our voids.
\citet{Ceccarelli2013} classified voids into those with and without surrounding overdense shells: large voids without shells tend to expand, while smaller voids surrounded by overdense shells are in the process of being crushed by the surrounding shell.
This is a useful division for understanding void dynamics, as predicted analytically by \citet{Sheth2004} \cs{and later studied in simulations \citep{Paz2013,Ceccarelli2013,Hamaus2014a} and data \citep{Ruiz2015}}.
Furthermore, this classification has been useful for predicting large-scale bulk flows of voids in both simulations \citep{Lambas2016} and data \citep{Ceccarelli2015}.
These works found that large voids are on average receding from each other, while small voids in overdense shells are approaching each other.

Most importantly, we have applied the algorithm to the DES-SV data and found a total of 87 voids over the redshift range $0.2<z<0.8$. Our $\sim 4\sigma$ detection of the weak gravitational lensing signal of these voids shows they are truly underdense in the matter field and hence not simply a product of Poisson noise, tracer density effects or any systematics in the data.
Assuming a model profile \citep{Hamaus2014a}, we find a best-fit central density of $\delta_c \sim -0.6$ and scale radius $r_s \sim R_v$.
Since $r_s$ is the void edge determined from lensing, and $R_v$ is the edge determined from the galaxy distribution, the best-fit lensing model shows consistency between the mass and galaxy distributions of voids.
Note however that the contours are broad and still allow for the possibility of $r_s \gtrsim R_v$.

Further applications of the same void finder will be explored in future DES data samples. Of particular interest is the study of the CMB cold imprint of voids (Kov\'acs et al. in prep), related to the properties and presence of Dark Energy through the integrated Sachs-Wolfe effect \citep{Granett2008,Cai2010,Cai2014,Hotchkiss2014}. 

The advances in this work towards finding voids in photometric surveys are also exciting in light of recent advances in void cosmology.
\citet{Clampitt2016} studied void-void and void-galaxy clustering and derived void bias using the spectroscopic SDSS luminous red galaxy (LRG) sample.
\cs{\citet{Hamaus2016} applied the Alcock-Paczynski test to void clustering statistics to put} $\sim$ 10\% constraints on $\Omega_m$ using voids identified using CMASS galaxies as tracers, a result that was anticipated in simulations by the same group \citep{Hamaus2014b, Hamaus2014, Hamaus2015}.
\citet{Kitaura2015} reported greater than $3\sigma$ evidence of the presence of baryonic acoustic oscillations (BAO) in void correlations, again using CMASS galaxies.
This impressive measurement was made possible by the new void finder presented in \citet{Zhao2015} and detailed studies with mock CMASS samples presented in \citet{Liang2015}.
While the CMASS sample from BOSS covers a very large area, it lacks a suitable background source sample for direct lensing measurements of void density profiles.
Upcoming photometric surveys, which will have many background sources available, will make the combination of void clustering and lensing over large volumes a reality.

In addition to constraining standard cosmological parameters, \cs{voids have been used to investigate alternative dark matter scenarios like warm dark matter \citep{Yang2015}, or the effects of neutrinos on void lensing \citep{Massara2015}. Especially numerous are the studies on void abundance \citep{Li2011, Clampitt2013, Cai2015, Zivick2015, Lam2015, Pollina2016} and lensing \citep{Cai2014, Barreira2015} as promising probes of alternatives to General Relativity (GR).}
In particular, \citet{Barreira2015} used simulations of Galileon gravity to show that the lensing signal of voids can be double that in GR.
Comparing to the SDSS void lensing results of \citet{Clampitt2014}, they showed that the size of the difference is comparable to current observational errors.
Furthermore, another recent development by \citet{Cautun2015} has shown that the signal-to-noise for void lensing can be increased by describing the void profile relative to the boundary rather than the center.
Such advances, combined with the increasing quality and volume of data from ongoing surveys, will bring modified gravity constraints using voids within reach. 
The algorithm in this work ensures that the statistical power of these new photometric datasets can be brought to bear on void measurements.

\section*{Acknowledgments}

This paper has gone through internal review by the DES collaboration. It has been assigned DES paper id DES-2016-0168 and FermiLab preprint number PUB-16-173-AE.

Funding for the DES Projects has been provided by the U.S. Department of Energy, the U.S. National Science Foundation, the Ministry of Science and Education of Spain, 
the Science and Technology Facilities Council of the United Kingdom, the Higher Education Funding Council for England, the National Center for Supercomputing 
Applications at the University of Illinois at Urbana-Champaign, the Kavli Institute of Cosmological Physics at the University of Chicago, 
the Center for Cosmology and Astro-Particle Physics at the Ohio State University,
the Mitchell Institute for Fundamental Physics and Astronomy at Texas A\&M University, Financiadora de Estudos e Projetos, 
Funda{\c c}{\~a}o Carlos Chagas Filho de Amparo {\`a} Pesquisa do Estado do Rio de Janeiro, Conselho Nacional de Desenvolvimento Cient{\'i}fico e Tecnol{\'o}gico and 
the Minist{\'e}rio da Ci{\^e}ncia, Tecnologia e Inova{\c c}{\~a}o, the Deutsche Forschungsgemeinschaft and the Collaborating Institutions in the Dark Energy Survey. 

The Collaborating Institutions are Argonne National Laboratory, the University of California at Santa Cruz, the University of Cambridge, Centro de Investigaciones Energ{\'e}ticas, 
Medioambientales y Tecnol{\'o}gicas-Madrid, the University of Chicago, University College London, the DES-Brazil Consortium, the University of Edinburgh, 
the Eidgen{\"o}ssische Technische Hochschule (ETH) Z{\"u}rich, 
Fermi National Accelerator Laboratory, the University of Illinois at Urbana-Champaign, the Institut de Ci{\`e}ncies de l'Espai (IEEC/CSIC), 
the Institut de F{\'i}sica d'Altes Energies, Lawrence Berkeley National Laboratory, the Ludwig-Maximilians Universit{\"a}t M{\"u}nchen and the associated Excellence Cluster Universe, 
the University of Michigan, the National Optical Astronomy Observatory, the University of Nottingham, The Ohio State University, the University of Pennsylvania, the University of Portsmouth, 
SLAC National Accelerator Laboratory, Stanford University, the University of Sussex, Texas A\&M University, and the OzDES Membership Consortium.

The DES data management system is supported by the National Science Foundation under Grant Number AST-1138766.
The DES participants from Spanish institutions are partially supported by MINECO under grants AYA2012-39559, ESP2013-48274, FPA2013-47986, and Centro de Excelencia Severo Ochoa SEV-2012-0234.
Research leading to these results has received funding from the European Research Council under the European Union's Seventh Framework Programme (FP7/2007-2013) including ERC grant agreements 
 240672, 291329, and 306478. Support for DG was provided by NASA through the Einstein Fellowship Program, grant PF5-160138.

We are grateful for the extraordinary contributions of our CTIO colleagues and the DECam Construction, Commissioning and Science Verification
teams in achieving the excellent instrument and telescope conditions that have made this work possible.  The success of this project also 
relies critically on the expertise and dedication of the DES Data Management group.

\bibliographystyle{mn2e}
\bibliography{/Users/csanchez/Dropbox/bibtex/library}

\begin{thebibliography}{92}
\expandafter\ifx\csname natexlab\endcsname\relax\def\natexlab#1{#1}\fi

\bibitem[{Aihara {et~al}\mbox{.}(2011)Aihara, {Allende Prieto}, An, Anderson,
  Aubourg, Balbinot, Beers, Berlind, Bickerton, Bizyaev, Blanton, Bochanski,
  Bolton, Bovy, Brandt, Brinkmann, Brown, Brownstein, Busca, Campbell, Carr,
  Chen, Chiappini, Comparat, Connolly, Cortes, Croft, Cuesta, da~Costa,
  Davenport, Dawson, Dhital, Ealet, Ebelke, Edmondson, Eisenstein, Escoffier,
  Esposito, Evans, Fan, {Femen{\'{\i}}a Castell{\'{a}}}, Font-Ribera,
  Frinchaboy, Ge, Gillespie, Gilmore, {Gonz{\'{a}}lez Hern{\'{a}}ndez}, Gott,
  Gould, Grebel, Gunn, Hamilton, Harding, Harris, Hawley, Hearty, Ho, Hogg,
  Holtzman, Honscheid, Inada, Ivans, Jiang, Johnson, Jordan, Jordan, Kazin,
  Kirkby, Klaene, Knapp, Kneib, Kochanek, Koesterke, Kollmeier, Kron, Lampeitl,
  Lang, {Le Goff}, Lee, Lin, Long, Loomis, Lucatello, Lundgren, Lupton, Ma,
  MacDonald, Mahadevan, Maia, Makler, Malanushenko, Malanushenko, Mandelbaum,
  Maraston, Margala, Masters, McBride, McGehee, McGreer, M{\'{e}}nard,
  Miralda-Escud{\'{e}}, Morrison, Mullally, Muna, Munn, Murayama, Myers,
  Naugle, {Fausti Neto}, Nguyen, Nichol, O'Connell, Ogando, Olmstead, Oravetz,
  Padmanabhan, Palanque-Delabrouille, Pan, Pandey, P{\^{a}}ris, Percival,
  Petitjean, Pfaffenberger, Pforr, Phleps, Pichon, Pieri, Prada, Price-Whelan,
  Raddick, Ramos, Reyl{\'{e}}, Rich, Richards, Rix, Robin, Rocha-Pinto,
  Rockosi, Roe, Rollinde, Ross, Ross, Rossetto, S{\'{a}}nchez, Sayres,
  Schlegel, Schlesinger, Schmidt, Schneider, Sheldon, Shu, Simmerer, Simmons,
  Sivarani, Snedden, Sobeck, Steinmetz, Strauss, Szalay, Tanaka, Thakar,
  Thomas, Tinker, Tofflemire, Tojeiro, Tremonti, Vandenberg, {Vargas
  Maga{\~{n}}a}, Verde, Vogt, Wake, Wang, Weaver, Weinberg, White, White,
  Yanny, Yasuda, Yeche, \& Zehavi}]{Aihara2011}
Aihara H. {et~al.}, 2011, Astrophys. J. Suppl. Ser., 193, 29

\bibitem[{Barreira {et~al}\mbox{.}(2015)Barreira, Cautun, Li, Baugh, \&
  Pascoli}]{Barreira2015}
Barreira A., Cautun M., Li B., Baugh C.~M., Pascoli S., 2015, J. Cosmol.
  Astropart. Phys., 2015, 028

\bibitem[{Becker {et~al}\mbox{.}(2015)Becker, Troxel, MacCrann, Krause, Eifler,
  Friedrich, Nicola, Refregier, Amara, Bacon, Bernstein, Bonnett, Bridle,
  Busha, Chang, Dodelson, Erickson, Evrard, Frieman, Gaztanaga, Gruen, Hartley,
  Jain, Jarvis, Kacprzak, Kirk, Kravtsov, Leistedt, Rykoff, Sabiu, Sanchez,
  Seo, Sheldon, Wechsler, Zuntz, Abbott, Abdalla, Allam, Armstrong, Banerji,
  Bauer, Benoit-Levy, Bertin, Brooks, Buckley-Geer, Burke, Capozzi, Rosell,
  Kind, Carretero, Castander, Crocce, Cunha, D'Andrea, da~Costa, DePoy, Desai,
  Diehl, Dietrich, Doel, Neto, Fernandez, Finley, Flaugher, Fosalba, Gerdes,
  Gruendl, Gutierrez, Honscheid, James, Kuehn, Kuropatkin, Lahav, Li, Lima,
  Maia, March, Martini, Melchior, Miller, Miquel, Mohr, Nichol, Nord, Ogando,
  Plazas, Reil, Romer, Roodman, Sako, Sanchez, Scarpine, Schubnell,
  Sevilla-Noarbe, Smith, Soares-Santos, Sobreira, Suchyta, Swanson, Tarle,
  Thaler, Thomas, Vikram, Walker, \& Collaboration}]{Becker2015}
Becker M.~R. {et~al.}, 2015, arXiv:1507.05598

\bibitem[{Behroozi, Wechsler \& Wu(2013)Behroozi, Wechsler, \&
  Wu}]{Behroozi2013}
Behroozi P.~S., Wechsler R.~H., Wu H.-Y., 2013, Astrophys. J., 762, 109

\bibitem[{Betancort-Rijo {et~al}\mbox{.}(2009)Betancort-Rijo, Patiri, Prada, \&
  Romano}]{Betancort-Rijo2009}
Betancort-Rijo J., Patiri S.~G., Prada F., Romano A.~E., 2009, Mon. Not. R.
  Astron. Soc., 400, 1835

\bibitem[{Bonnett {et~al}\mbox{.}(2015)Bonnett, Troxel, Hartley, Amara,
  Leistedt, Becker, Bernstein, Bridle, Bruderer, Busha, Kind, Childress,
  Castander, Chang, Crocce, Davis, Eifler, Frieman, Gangkofner, Gaztanaga,
  Glazebrook, Gruen, Kacprzak, King, Kwan, Lahav, Lewis, Lidman, Lin, MacCrann,
  Miquel, O'Neill, Palmese, Peiris, Refregier, Rozo, Rykoff, Sadeh,
  S{\'{a}}nchez, Sheldon, Uddin, Wechsler, Zuntz, Abbott, Abdalla, Allam,
  Armstrong, Banerji, Bauer, Benoit-L{\'{e}}vy, Bertin, Brooks, Buckley-Geer,
  Burke, Capozzi, Rosell, Carretero, Cunha, D'Andrea, da~Costa, DePoy, Desai,
  Diehl, Dietrich, Doel, Neto, Fernandez, Flaugher, Fosalba, Gerdes, Gruendl,
  Honscheid, Jain, James, Jarvis, Kim, Kuehn, Kuropatkin, Li, Lima, Maia,
  March, Marshall, Martini, Melchior, Miller, Neilsen, Nichol, Nord, Ogando,
  Plazas, Reil, Romer, Roodman, Sako, Sanchez, Santiago, Smith, Soares-Santos,
  Sobreira, Suchyta, Swanson, Tarle, Thaler, Thomas, Vikram, \&
  Walker}]{Bonnett2015}
Bonnett C. {et~al.}, 2015, arXiv:1507.05909

\bibitem[{Bos {et~al}\mbox{.}(2012)Bos, van~de Weygaert, Dolag, \&
  Pettorino}]{Bos2012}
Bos E. G.~P., van~de Weygaert R., Dolag K., Pettorino V., 2012, Mon. Not. R.
  Astron. Soc., 426, 440

\bibitem[{Bruzual \& Charlot(2003)}]{Bruzual2003}
Bruzual G., Charlot S., 2003, Mon. Not. R. Astron. Soc., 344, 1000

\bibitem[{Busha {et~al}\mbox{.}(2013)Busha, Wechsler, Becker, Erickson, \&
  Evrard}]{Busha2013}
Busha M.~T., Wechsler R.~H., Becker M.~R., Erickson B., Evrard A.~E., 2013, Am.
  Astron. Soc. AAS Meet. {\#}221, id.341.07

\bibitem[{Cai {et~al}\mbox{.}(2010)Cai, Cole, Jenkins, \& Frenk}]{Cai2010}
Cai Y.-C., Cole S., Jenkins A., Frenk C.~S., 2010, Mon. Not. R. Astron. Soc.,
  407, 201

\bibitem[{Cai, Padilla \& Li(2014)Cai, Padilla, \& Li}]{Cai2014}
Cai Y.-C., Padilla N., Li B., 2014, arXiv:1410.8355

\bibitem[{Cai, Padilla \& Li(2015)Cai, Padilla, \& Li}]{Cai2015}
Cai Y.-C., Padilla N., Li B., 2015, Mon. Not. R. Astron. Soc., 451, 1036

\bibitem[{Cautun, Cai \& Frenk(2015)Cautun, Cai, \& Frenk}]{Cautun2015}
Cautun M., Cai Y.-C., Frenk C.~S., 2015, arXiv:1509.00010

\bibitem[{Ceccarelli {et~al}\mbox{.}(2006)Ceccarelli, Padilla, Valotto, \&
  Lambas}]{Ceccarelli2006}
Ceccarelli L., Padilla N.~D., Valotto C., Lambas D.~G., 2006, Mon. Not. R.
  Astron. Soc., 373, 1440

\bibitem[{Ceccarelli {et~al}\mbox{.}(2013)Ceccarelli, Paz, Lares, Padilla, \&
  Lambas}]{Ceccarelli2013}
Ceccarelli L., Paz D., Lares M., Padilla N., Lambas D.~G., 2013, Mon. Not. R.
  Astron. Soc., 434, 1435

\bibitem[{Ceccarelli {et~al}\mbox{.}(2015)Ceccarelli, Ruiz, Lares, Paz,
  Maldonado, Luparello, \& Lambas}]{Ceccarelli2015}
Ceccarelli L., Ruiz A.~N., Lares M., Paz D.~J., Maldonado V.~E., Luparello
  H.~E., Lambas D.~G., 2015, arXiv:1511.06741

\bibitem[{Chan, Hamaus \& Desjacques(2014)Chan, Hamaus, \&
  Desjacques}]{Chan2014}
Chan K.~C., Hamaus N., Desjacques V., 2014, Phys. Rev. D, 90, 103521

\bibitem[{Chang {et~al}\mbox{.}(2015)Chang, Vikram, Jain, Bacon, Amara, Becker,
  Bernstein, Bonnett, Bridle, Brout, Busha, Frieman, Gaztanaga, Hartley,
  Jarvis, Kacprzak, Kov{\'{a}}cs, Lahav, Lin, Melchior, Peiris, Rozo, Rykoff,
  S{\'{a}}nchez, Sheldon, Troxel, Wechsler, Zuntz, Abbott, Abdalla, Allam,
  Annis, Bauer, Benoit-L{\'{e}}vy, Brooks, Buckley-Geer, Burke, Capozzi,
  {Carnero Rosell}, {Carrasco Kind}, Castander, Crocce, D’Andrea, Desai,
  Diehl, Dietrich, Doel, Eifler, Evrard, {Fausti Neto}, Flaugher, Fosalba,
  Gruen, Gruendl, Gutierrez, Honscheid, James, Kent, Kuehn, Kuropatkin, Maia,
  March, Martini, Merritt, Miller, Miquel, Neilsen, Nichol, Ogando, Plazas,
  Romer, Roodman, Sako, Sanchez, Sevilla, Smith, Soares-Santos, Sobreira,
  Suchyta, Tarle, Thaler, Thomas, Tucker, \& Walker}]{Chang2015}
Chang C. {et~al.}, 2015, Phys. Rev. Lett., 115, 051301

\bibitem[{Clampitt, Cai \& Li(2013)Clampitt, Cai, \& Li}]{Clampitt2013}
Clampitt J., Cai Y.-C., Li B., 2013, Mon. Not. R. Astron. Soc., 431, 749

\bibitem[{Clampitt \& Jain(2015)}]{Clampitt2014}
Clampitt J., Jain B., 2015, Mon. Not. R. Astron. Soc., 454, 3357

\bibitem[{Clampitt, Jain \& S{\'{a}}nchez(2016)Clampitt, Jain, \&
  S{\'{a}}nchez}]{Clampitt2016}
Clampitt J., Jain B., S{\'{a}}nchez C., 2016, Mon. Not. R. Astron. Soc., 456,
  4425

\bibitem[{Clampitt {et~al}\mbox{.}(2016)Clampitt, S{\'{a}}nchez, Kwan, Krause,
  MacCrann, Park, Troxel, Jain, Rozo, Rykoff, Wechsler, Blazek, Bonnett,
  Crocce, Fang, Gaztanaga, Gruen, Jarvis, Miquel, Prat, Ross, Sheldon, Zuntz,
  Abbott, Abdalla, Armstrong, Becker, Benoit-L{\'{e}}vy, Bernstein, Bertin,
  Brooks, Burke, Rosell, Kind, Cunha, D'Andrea, da~Costa, Desai, Diehl,
  Dietrich, Doel, Estrada, Evrard, Neto, Flaugher, Fosalba, Frieman, Gruendl,
  Honscheid, James, Kuehn, Kuropatkin, Lahav, Lima, March, Marshall, Martini,
  Melchior, Mohr, Nichol, Nord, Plazas, Romer, Sanchez, Scarpine, Schubnell,
  Sevilla-Noarbe, Smith, Soares-Santos, Sobreira, Suchyta, Swanson, Tarle,
  Thomas, Vikram, \& Walker}]{Clampitt2016a}
Clampitt J. {et~al.}, 2016, arXiv:1603.05790

\bibitem[{Colberg {et~al}\mbox{.}(2008)Colberg, Pearce, Foster, Platen,
  Brunino, Neyrinck, Basilakos, Fairall, Feldman, Gottl{\"{o}}ber, Hahn, Hoyle,
  M{\"{u}}ller, Nelson, Plionis, Porciani, Shandarin, Vogeley, \& van~de
  Weygaert}]{Colberg2008}
Colberg J.~M. {et~al.}, 2008, Mon. Not. R. Astron. Soc., 387, 933

\bibitem[{Colberg {et~al}\mbox{.}(2005)Colberg, Sheth, Diaferio, Gao, \&
  Yoshida}]{Colberg2005}
Colberg J.~M., Sheth R.~K., Diaferio A., Gao L., Yoshida N., 2005, Mon. Not. R.
  Astron. Soc., 360, 216

\bibitem[{Colless {et~al}\mbox{.}(2001)Colless, Dalton, Maddox, Sutherland,
  Norberg, Cole, Bland-Hawthorn, Bridges, Cannon, Collins, Couch, Cross,
  Deeley, {De Propris}, Driver, Efstathiou, Ellis, Frenk, Glazebrook, Jackson,
  Lahav, Lewis, Lumsden, Madgwick, Peacock, Peterson, Price, Seaborne, \&
  Taylor}]{Colless2001}
Colless M. {et~al.}, 2001, Mon. Not. R. Astron. Soc., 328, 1039

\bibitem[{Conroy, Wechsler \& Kravtsov(2006)Conroy, Wechsler, \&
  Kravtsov}]{Conroy2006}
Conroy C., Wechsler R.~H., Kravtsov A.~V., 2006, Astrophys. J., 647, 201

\bibitem[{Crocce {et~al}\mbox{.}(2016)Crocce, Carretero, Bauer, Ross,
  Sevilla-Noarbe, Giannantonio, Sobreira, Sanchez, Gaztanaga, Kind,
  S{\'{a}}nchez, Bonnett, Benoit-L{\'{e}}vy, Brunner, Rosell, Cawthon, Fosalba,
  Hartley, Kim, Leistedt, Miquel, Peiris, Percival, Rosenfeld, Rykoff,
  S{\'{a}}nchez, Abbott, Abdalla, Allam, Banerji, Bernstein, Bertin, Brooks,
  Buckley-Geer, Burke, Capozzi, Castander, Cunha, D'Andrea, da~Costa, Desai,
  Diehl, Eifler, Evrard, Neto, Fernandez, Finley, Flaugher, Frieman, Gerdes,
  Gruen, Gruendl, Gutierrez, Honscheid, James, Kuehn, Kuropatkin, Lahav, Li,
  Lima, Maia, March, Marshall, Martini, Melchior, Miller, Neilsen, Nichol,
  Nord, Ogando, Plazas, Romer, Sako, Santiago, Schubnell, Smith, Soares-Santos,
  Suchyta, Swanson, Tarle, Thaler, Thomas, Vikram, Walker, Wechsler, Weller, \&
  Zuntz}]{Crocce2016}
Crocce M. {et~al.}, 2016, Mon. Not. R. Astron. Soc., 455, 4301

\bibitem[{Crocce {et~al}\mbox{.}(2011)Crocce, Gazta{\~{n}}aga, Cabr{\'{e}},
  Carnero, \& S{\'{a}}nchez}]{Crocce2011}
Crocce M., Gazta{\~{n}}aga E., Cabr{\'{e}} A., Carnero A., S{\'{a}}nchez E.,
  2011, Mon. Not. R. Astron. Soc., 417, 2577

\bibitem[{Crocce, Pueblas \& Scoccimarro(2006)Crocce, Pueblas, \&
  Scoccimarro}]{Crocce2006}
Crocce M., Pueblas S., Scoccimarro R., 2006, Mon. Not. R. Astron. Soc., 373,
  369

\bibitem[{{Dark Energy Survey Collaboration} {et~al}\mbox{.}(2016){Dark Energy
  Survey Collaboration}, Abbott, Abdalla, Allam, Aleksic, Amara, Bacon,
  Balbinot, Banerji, Bechtol, Benoit-Levy, Bernstein, Bertin, Blazek, Dodelson,
  Bonnett, Brooks, Bridle, Brunner, Buckley-Geer, Burke, Capozzi, Caminha,
  Carlsen, Carnero-Rosell, Carollo, Carrasco-Kind, Carretero, Castander,
  Clerkin, Collett, Conselice, Crocce, Cunha, D'Andrea, da~Costa, Davis, Desai,
  Diehl, Dietrich, Doel, Drlica-Wagner, Etherington, Estrada, Evrard, Fabbri,
  Finley, Flaugher, Fosalba, Foley, Frieman, Garcia-Bellido, Gaztanaga, Gerdes,
  Giannantonio, Goldstein, Gruen, Gruendl, Guarnieri, Gutierrez, Hartley,
  Honscheid, Jain, James, Jeltema, Jouvel, Kessler, King, Kirk, Kron, Kuehn,
  Kuropatkin, Lahav, Li, Lima, Lin, Maia, Makler, Manera, Maraston, Marshall,
  Martini, McMahon, Melchior, Merson, Miller, Miquel, Mohr, Morice-Atkinson,
  Naidoo, Neilsen, Nichol, Nord, Ogando, Ostrovski, Palmese, Papadopoulos,
  Peiris, Peoples, Plazas, Percival, Reed, Romer, Roodman, Ross, Rozo, Rykoff,
  Sadeh, Sako, Sanchez, Sanchez, Santiago, Scarpine, Schubnell, Sevilla-Noarbe,
  Sheldon, Smith, Smith, Soares-Santos, Sobreira, Soumagnac, Suchyta, Sullivan,
  Swanson, Tarle, Thaler, Thomas, Thomas, Tucker, Vieira, Vikram, Walker,
  Wechsler, Wester, Weller, Whiteway, Wilcox, Yanny, Zhang, \&
  Zuntz}]{DarkEnergySurveyCollaboration2016}
{Dark Energy Survey Collaboration} {et~al.}, 2016, Mon. Not. R. Astron. Soc.,
  stw641

\bibitem[{Dawson {et~al}\mbox{.}(2013)Dawson, Schlegel, Ahn, Anderson, Aubourg,
  Bailey, Barkhouser, Bautista, Beifiori, Berlind, Bhardwaj, Bizyaev, Blake,
  Blanton, Blomqvist, Bolton, Borde, Bovy, Brandt, Brewington, Brinkmann,
  Brown, Brownstein, Bundy, Busca, Carithers, Carnero, Carr, Chen, Comparat,
  Connolly, Cope, Croft, Cuesta, da~Costa, Davenport, Delubac, de~Putter,
  Dhital, Ealet, Ebelke, Eisenstein, Escoffier, Fan, {Filiz Ak}, Finley,
  Font-Ribera, G{\'{e}}nova-Santos, Gunn, Guo, Haggard, Hall, Hamilton, Harris,
  Harris, Ho, Hogg, Holder, Honscheid, Huehnerhoff, Jordan, Jordan, Kauffmann,
  Kazin, Kirkby, Klaene, Kneib, {Le Goff}, Lee, Long, Loomis, Lundgren, Lupton,
  Maia, Makler, Malanushenko, Malanushenko, Mandelbaum, Manera, Maraston,
  Margala, Masters, McBride, McDonald, McGreer, McMahon, Mena,
  Miralda-Escud{\'{e}}, Montero-Dorta, Montesano, Muna, Myers, Naugle, Nichol,
  Noterdaeme, Nuza, Olmstead, Oravetz, Oravetz, Owen, Padmanabhan,
  Palanque-Delabrouille, Pan, Parejko, P{\^{a}}ris, Percival,
  P{\'{e}}rez-Fournon, P{\'{e}}rez-R{\`{a}}fols, Petitjean, Pfaffenberger,
  Pforr, Pieri, Prada, Price-Whelan, Raddick, Rebolo, Rich, Richards, Rockosi,
  Roe, Ross, Ross, Rossi, Rubi{\~{n}}o-Martin, Samushia, S{\'{a}}nchez, Sayres,
  Schmidt, Schneider, Sc{\'{o}}ccola, Seo, Shelden, Sheldon, Shen, Shu, Slosar,
  Smee, Snedden, Stauffer, Steele, Strauss, Streblyanska, Suzuki, Swanson, Tal,
  Tanaka, Thomas, Tinker, Tojeiro, Tremonti, {Vargas Maga{\~{n}}a}, Verde,
  Viel, Wake, Watson, Weaver, Weinberg, Weiner, West, White, Wood-Vasey, Yeche,
  Zehavi, Zhao, \& Zheng}]{Dawson2013}
Dawson K.~S. {et~al.}, 2013, Astron. J., 145, 10

\bibitem[{de~Jong {et~al}\mbox{.}(2013)de~Jong, Kuijken, Applegate, Begeman,
  Belikov, \& al}]{Jong2013}
de~Jong J. T.~A., Kuijken K., Applegate D., Begeman K., Belikov A., al E.,
  2013, ESO Messenger, 154, 44

\bibitem[{Diehl {et~al}\mbox{.}(2014)Diehl, Abbott, Annis, Armstrong, Baruah,
  Bermeo, Bernstein, Beynon, Bruderer, Buckley-Geer, Campbell, Capozzi, Carter,
  Casas, Clerkin, Covarrubias, Cuhna, D'Andrea, da~Costa, Das, DePoy, Dietrich,
  Drlica-Wagner, Elliott, Eifler, Estrada, Etherington, Flaugher, Frieman,
  {Fausti Neto}, Gelman, Gerdes, Gruen, Gruendl, Hao, Head, Helsby, Hoffman,
  Honscheid, James, Johnson, Kacprzac, Katsaros, Kennedy, Kent, Kessler, Kim,
  Krause, Kron, Kuhlmann, Kunder, Li, Lin, MacCrann, March, Marshall, Neilsen,
  Nugent, Martini, Melchior, Menanteau, Nichol, Nord, Ogando, Old,
  Papadopoulos, Patton, Petravick, Plazas, Poulton, Pujol, Reil, Rigby, Romer,
  Roodman, Rooney, {Sanchez Alvaro}, Serrano, Sheldon, Smith, Smith,
  Soares-Santos, Soumagnac, Spinka, Suchyta, Tucker, Walker, Wester, Wiesner,
  Wilcox, Williams, Yanny, \& Zhang}]{Diehl2014}
Diehl H.~T. {et~al.}, 2014, in Proc. SPIE, Peck A.~B., Benn C.~R., Seaman
  R.~L., eds., p. 91490V

\bibitem[{Drinkwater {et~al}\mbox{.}(2010)Drinkwater, Jurek, Blake, Woods,
  Pimbblet, Glazebrook, Sharp, Pracy, Brough, Colless, Couch, Croom, Davis,
  Forbes, Forster, Gilbank, Gladders, Jelliffe, Jones, Li, Madore, Martin,
  Poole, Small, Wisnioski, Wyder, \& Yee}]{Drinkwater2010}
Drinkwater M.~J. {et~al.}, 2010, Mon. Not. R. Astron. Soc., 401, 1429

\bibitem[{Elyiv {et~al}\mbox{.}(2015)Elyiv, Marulli, Pollina, Baldi, Branchini,
  Cimatti, \& Moscardini}]{Elyiv2015}
Elyiv A., Marulli F., Pollina G., Baldi M., Branchini E., Cimatti A.,
  Moscardini L., 2015, Mon. Not. R. Astron. Soc., 448, 642

\bibitem[{Flaugher(2005)}]{Flaugher2005}
Flaugher B., 2005, Int. J. Mod. Phys., A20, 3121

\bibitem[{Flaugher {et~al}\mbox{.}(2015)Flaugher, Diehl, Honscheid, Abbott,
  Alvarez, Angstadt, Annis, Antonik, Ballester, Beaufore, Bernstein, Bernstein,
  Bigelow, Bonati, Boprie, Brooks, Buckley-Geer, Campa, Cardiel-Sas, Castander,
  Castilla, Cease, Cela-Ruiz, Chappa, Chi, Cooper, da~Costa, Dede, Derylo,
  DePoy, de~Vicente, Doel, Drlica-Wagner, Eiting, Elliott, Emes, Estrada,
  {Fausti Neto}, Finley, Flores, Frieman, Gerdes, Gladders, Gregory, Gutierrez,
  Hao, Holland, Holm, Huffman, Jackson, James, Jonas, Karcher, Karliner, Kent,
  Kessler, Kozlovsky, Kron, Kubik, Kuehn, Kuhlmann, Kuk, Lahav, Lathrop, Lee,
  Levi, Lewis, Li, Mandrichenko, Marshall, Martinez, Merritt, Miquel,
  Mu{\~{n}}oz, Neilsen, Nichol, Nord, Ogando, Olsen, Palaio, Patton, Peoples,
  Plazas, Rauch, Reil, Rheault, Roe, Rogers, Roodman, Sanchez, Scarpine,
  Schindler, Schmidt, Schmitt, Schubnell, Schultz, Schurter, Scott, Serrano,
  Shaw, Smith, Soares-Santos, Stefanik, Stuermer, Suchyta, Sypniewski, Tarle,
  Thaler, Tighe, Tran, Tucker, Walker, Wang, Watson, Weaverdyck, Wester, Woods,
  \& Yanny}]{Flaugher2015}
Flaugher B. {et~al.}, 2015, Astron. J., 150, 150

\bibitem[{Gorski {et~al}\mbox{.}(2005)Gorski, Hivon, Banday, Wandelt, Hansen,
  Reinecke, \& Bartelmann}]{Gorski2005}
Gorski K.~M., Hivon E., Banday A.~J., Wandelt B.~D., Hansen F.~K., Reinecke M.,
  Bartelmann M., 2005, Astrophys. J., 622, 759

\bibitem[{Granett, Neyrinck \& Szapudi(2008)Granett, Neyrinck, \&
  Szapudi}]{Granett2008}
Granett B.~R., Neyrinck M.~C., Szapudi I., 2008, Astrophys. J., 683, L99

\bibitem[{Gruen {et~al}\mbox{.}(2016)Gruen, Friedrich, Amara, Bacon, Bonnett,
  Hartley, Jain, Jarvis, Kacprzak, Krause, Mana, Rozo, Rykoff, Seitz, Sheldon,
  Troxel, Vikram, Abbott, Abdalla, Allam, Armstrong, Banerji, Bauer, Becker,
  Benoit-L{\'{e}}vy, Bernstein, Bernstein, Bertin, Bridle, Brooks,
  Buckley-Geer, Burke, Capozzi, {Carnero Rosell}, {Carrasco Kind}, Carretero,
  Crocce, Cunha, D'Andrea, da~Costa, DePoy, Desai, Diehl, Dietrich, Doel,
  Eifler, Neto, Fernandez, Flaugher, Fosalba, Frieman, Gerdes, Gruendl,
  Gutierrez, Honscheid, James, Kuehn, Kuropatkin, Lahav, Li, Lima, Maia, March,
  Martini, Melchior, Miller, Miquel, Mohr, Nord, Ogando, Plazas, Reil, Romer,
  Roodman, Sako, Sanchez, Scarpine, Schubnell, Sevilla-Noarbe, Smith,
  Soares-Santos, Sobreira, Suchyta, Swanson, Tarle, Thaler, Thomas, Walker,
  Wechsler, Weller, Zhang, \& Zuntz}]{Gruen2015}
Gruen D. {et~al.}, 2016, Mon. Not. R. Astron. Soc., 455, 3367

\bibitem[{Hamaus {et~al}\mbox{.}(2016)Hamaus, Pisani, Sutter, Lavaux,
  Escoffier, Wandelt, \& Weller}]{Hamaus2016}
Hamaus N., Pisani A., Sutter P.~M., Lavaux G., Escoffier S., Wandelt B.~D.,
  Weller J., 2016, arXiv:1602.01784

\bibitem[{Hamaus {et~al}\mbox{.}(2014{\natexlab{a}})Hamaus, Sutter, Lavaux, \&
  Wandelt}]{Hamaus2014b}
Hamaus N., Sutter P., Lavaux G., Wandelt B.~D., 2014{\natexlab{a}}, J. Cosmol.
  Astropart. Phys., 2014, 013

\bibitem[{Hamaus {et~al}\mbox{.}(2015)Hamaus, Sutter, Lavaux, \&
  Wandelt}]{Hamaus2015}
Hamaus N., Sutter P., Lavaux G., Wandelt B.~D., 2015, J. Cosmol. Astropart.
  Phys., 2015, 036

\bibitem[{Hamaus, Sutter \& Wandelt(2014)Hamaus, Sutter, \&
  Wandelt}]{Hamaus2014a}
Hamaus N., Sutter P.~M., Wandelt B.~D., 2014, Phys. Rev. Lett., 112, 251302

\bibitem[{Hamaus {et~al}\mbox{.}(2014{\natexlab{b}})Hamaus, Wandelt, Sutter,
  Lavaux, \& Warren}]{Hamaus2014}
Hamaus N., Wandelt B.~D., Sutter P.~M., Lavaux G., Warren M.~S.,
  2014{\natexlab{b}}, Phys. Rev. Lett., 112, 041304

\bibitem[{Hotchkiss {et~al}\mbox{.}(2014)Hotchkiss, Nadathur, Gottlober, Iliev,
  Knebe, Watson, \& Yepes}]{Hotchkiss2014}
Hotchkiss S., Nadathur S., Gottlober S., Iliev I.~T., Knebe A., Watson W.~A.,
  Yepes G., 2014, Mon. Not. R. Astron. Soc., 446, 1321

\bibitem[{Hoyle \& Vogeley(2002)}]{Hoyle2002}
Hoyle F., Vogeley M.~S., 2002, Astrophys. J., 566, 641

\bibitem[{Jarvis {et~al}\mbox{.}(2015)Jarvis, Sheldon, Zuntz, Kacprzak, Bridle,
  Amara, Armstrong, Becker, Bernstein, Bonnett, Chang, Das, Dietrich,
  Drlica-Wagner, Eifler, Gangkofner, Gruen, Hirsch, Huff, Jain, Kent, Kirk,
  MacCrann, Melchior, Plazas, Refregier, Rowe, Rykoff, Samuroff, S{\'{a}}nchez,
  Suchyta, Troxel, Vikram, Abbott, Abdalla, Allam, Annis, Benoit-L{\'{e}}vy,
  Bertin, Brooks, Buckley-Geer, Burke, Capozzi, Rosell, Kind, Carretero,
  Castander, Crocce, Cunha, D'Andrea, da~Costa, DePoy, Desai, Diehl, Doel,
  Neto, Flaugher, Fosalba, Frieman, Gaztanaga, Gerdes, Gruendl, Gutierrez,
  Honscheid, James, Kuehn, Kuropatkin, Lahav, Li, Lima, March, Martini, Miquel,
  Mohr, Neilsen, Nord, Ogando, Reil, Romer, Roodman, Sako, Sanchez, Scarpine,
  Schubnell, Sevilla-Noarbe, Smith, Soares-Santos, Sobreira, Swanson, Tarle,
  Thaler, Thomas, Walker, \& Wechsler}]{Jarvis2015}
Jarvis M. {et~al.}, 2015, arXiv:1507.05603

\bibitem[{Jennings, Li \& Hu(2013)Jennings, Li, \& Hu}]{Jennings2013}
Jennings E., Li Y., Hu W., 2013, Mon. Not. R. Astron. Soc., 434, 2167

\bibitem[{Kaiser, Tonry \& Luppino(2000)Kaiser, Tonry, \& Luppino}]{Kaiser2000}
Kaiser N., Tonry J.~L., Luppino G.~A., 2000, Publ. Astron. Soc. Pacific, 112,
  768

\bibitem[{Kitaura {et~al}\mbox{.}(2015)Kitaura, Chuang, Liang, Zhao, Tao,
  Rodriguez-Torres, Eisenstein, Gil-Marin, Kneib, McBride, Percival, Ross,
  Sanchez, Tinker, Tojeiro, Vargas-Magana, \& Zhao}]{Kitaura2015}
Kitaura F.-S. {et~al.}, 2015, arXiv:1511.04405

\bibitem[{Kwan {et~al}\mbox{.}(2016)Kwan, Sanchez, Clampitt, Blazek, Crocce,
  Jain, Zuntz, Amara, Becker, Bernstein, Bonnett, DeRose, Dodelson, Eifler,
  Gaztanaga, Giannantonio, Gruen, Hartley, Kacprzak, Kirk, Krause, MacCrann,
  Miquel, Park, Ross, Rozo, Rykoff, Sheldon, Troxel, Wechsler, Abbott, Abdalla,
  Allam, Benoit-L{\'{e}}vy, Brooks, Burke, Rosell, Kind, Cunha, D'Andrea,
  da~Costa, Desai, Diehl, Dietrich, Doel, Evrard, Fernandez, Finley, Flaugher,
  Fosalba, Frieman, Gerdes, Gruendl, Gutierrez, Honscheid, James, Jarvis,
  Kuehn, Lahav, Lima, Maia, Marshall, Martini, Melchior, Mohr, Nichol, Nord,
  Plazas, Reil, Romer, Roodman, Sanchez, Scarpine, Sevilla, Smith,
  Soares-Santos, Sobreira, Suchyta, Swanson, Tarle, Thomas, Vikram, \&
  Walker}]{Kwan2016}
Kwan J. {et~al.}, 2016, arXiv:1604.07871

\bibitem[{Lam {et~al}\mbox{.}(2015)Lam, Clampitt, Cai, \& Li}]{Lam2015}
Lam T.~Y., Clampitt J., Cai Y.-C., Li B., 2015, Mon. Not. R. Astron. Soc., 450,
  3319

\bibitem[{Lambas {et~al}\mbox{.}(2016)Lambas, Lares, Ceccarelli, Ruiz, Paz,
  Maldonado, \& Luparello}]{Lambas2016}
Lambas D.~G., Lares M., Ceccarelli L., Ruiz A.~N., Paz D.~J., Maldonado V.~E.,
  Luparello H.~E., 2016, Mon. Not. R. Astron. Soc. Lett., 455, L99

\bibitem[{Lavaux \& Wandelt(2010)}]{Lavaux2010}
Lavaux G., Wandelt B.~D., 2010, Mon. Not. R. Astron. Soc., 403, 1392

\bibitem[{Lavaux \& Wandelt(2012)}]{Lavaux2012}
Lavaux G., Wandelt B.~D., 2012, Astrophys. J., 754, 109

\bibitem[{{Le F{\`{e}}vre} {et~al}\mbox{.}(2005){Le F{\`{e}}vre}, Vettolani,
  Garilli, Tresse, Bottini, {Le Brun}, Maccagni, Picat, Scaramella, Scodeggio,
  Zanichelli, Adami, Arnaboldi, Arnouts, Bardelli, Bolzonella, Cappi, Charlot,
  Ciliegi, Contini, Foucaud, Franzetti, Gavignaud, Guzzo, Ilbert, Iovino,
  McCracken, Marano, Marinoni, Mathez, Mazure, Meneux, Merighi, Paltani,
  Pell{\`{o}}, Pollo, Pozzetti, Radovich, Zamorani, Zucca, Bondi, Bongiorno,
  Busarello, Lamareille, Mellier, Merluzzi, Ripepi, \& Rizzo}]{LeFevre2005}
{Le F{\`{e}}vre} O. {et~al.}, 2005, Astron. Astrophys., 439, 845

\bibitem[{Lee \& Park(2009)}]{Lee2009}
Lee J., Park D., 2009, Astrophys. J., 696, L10

\bibitem[{Leistedt {et~al}\mbox{.}(2015)Leistedt, Peiris, Elsner,
  Benoit-L{\'{e}}vy, Amara, Bauer, Becker, Bonnett, Bruderer, Busha, Kind,
  Chang, Crocce, da~Costa, Gaztanaga, Huff, Lahav, Palmese, Percival,
  Refregier, Ross, Rozo, Rykoff, S{\'{a}}nchez, Sadeh, Sevilla-Noarbe,
  Sobreira, Suchyta, Swanson, Wechsler, Abdalla, Allam, Banerji, Bernstein,
  Bernstein, Bertin, Bridle, Brooks, Buckley-Geer, Burke, Capozzi, Rosell,
  Carretero, Cunha, D'Andrea, DePoy, Desai, Diehl, Doel, Eifler, Evrard, Neto,
  Flaugher, Fosalba, Frieman, Gerdes, Gruen, Gruendl, Gutierrez, Honscheid,
  James, Jarvis, Kent, Kuehn, Kuropatkin, Li, Lima, Maia, March, Marshall,
  Martini, Melchior, Miller, Miquel, Nichol, Nord, Ogando, Plazas, Reil, Romer,
  Roodman, Sanchez, Santiago, Scarpine, Schubnell, Smith, Soares-Santos, Tarle,
  Thaler, Thomas, Vikram, Walker, Wester, Zhang, \& Zuntz}]{Leistedt2015}
Leistedt B. {et~al.}, 2015, arXiv:1507.05647

\bibitem[{Lewis \& Bridle(2002)}]{Lewis2002}
Lewis A., Bridle S., 2002, Phys. Rev. D, 66, 103511

\bibitem[{Li(2011)}]{Li2011}
Li B., 2011, Mon. Not. R. Astron. Soc., 411, 2615

\bibitem[{Liang {et~al}\mbox{.}(2015)Liang, Zhao, Chuang, Kitaura, \&
  Tao}]{Liang2015}
Liang Y., Zhao C., Chuang C.-H., Kitaura F.-S., Tao C., 2015, arXiv:1511.04391

\bibitem[{Mao {et~al}\mbox{.}(2016)Mao, Berlind, Scherrer, Neyrinck,
  Scoccimarro, Tinker, McBride, \& Schneider}]{Mao2016}
Mao Q., Berlind A.~A., Scherrer R.~J., Neyrinck M.~C., Scoccimarro R., Tinker
  J.~L., McBride C.~K., Schneider D.~P., 2016, arXiv:1602.06306

\bibitem[{Massara {et~al}\mbox{.}(2015)Massara, Villaescusa-Navarro, Viel, \&
  Sutter}]{Massara2015}
Massara E., Villaescusa-Navarro F., Viel M., Sutter P., 2015, J. Cosmol.
  Astropart. Phys., 2015, 018

\bibitem[{Melchior {et~al}\mbox{.}(2014)Melchior, Sutter, Sheldon, Krause, \&
  Wandelt}]{Melchior2014}
Melchior P., Sutter P.~M., Sheldon E.~S., Krause E., Wandelt B.~D., 2014, Mon.
  Not. R. Astron. Soc., 440, 2922

\bibitem[{Nadathur {et~al}\mbox{.}(2015)Nadathur, Hotchkiss, Diego, Iliev,
  Gottlober, Watson, \& Yepes}]{Nadathur2015}
Nadathur S., Hotchkiss S., Diego J.~M., Iliev I.~T., Gottlober S., Watson
  W.~A., Yepes G., 2015, Mon. Not. R. Astron. Soc., 449, 3997

\bibitem[{Neyrinck(2008)}]{Neyrinck2008}
Neyrinck M.~C., 2008, Mon. Not. R. Astron. Soc., 386, 2101

\bibitem[{Norberg {et~al}\mbox{.}(2009)Norberg, Baugh, Gazta{\~{n}}aga, \&
  Croton}]{Norberg2009}
Norberg P., Baugh C.~M., Gazta{\~{n}}aga E., Croton D.~J., 2009, Mon. Not. R.
  Astron. Soc., 396, 19

\bibitem[{Padilla, Ceccarelli \& Lambas(2005)Padilla, Ceccarelli, \&
  Lambas}]{Padilla2005}
Padilla N.~D., Ceccarelli L., Lambas D.~G., 2005, Mon. Not. R. Astron. Soc.,
  363, 977

\bibitem[{Paz {et~al}\mbox{.}(2013)Paz, Lares, Ceccarelli, Padilla, \&
  Lambas}]{Paz2013}
Paz D., Lares M., Ceccarelli L., Padilla N., Lambas D.~G., 2013, Mon. Not. R.
  Astron. Soc., 436, 3480

\bibitem[{Platen, {Van De Weygaert} \& Jones(2007)Platen, {Van De Weygaert}, \&
  Jones}]{Platen2007}
Platen E., {Van De Weygaert} R., Jones B. J.~T., 2007, Mon. Not. R. Astron.
  Soc., 380, 551

\bibitem[{Pollina {et~al}\mbox{.}(2016)Pollina, Baldi, Marulli, \&
  Moscardini}]{Pollina2016}
Pollina G., Baldi M., Marulli F., Moscardini L., 2016, Mon. Not. R. Astron.
  Soc., 455, 3075

\bibitem[{Reddick {et~al}\mbox{.}(2013)Reddick, Wechsler, Tinker, \&
  Behroozi}]{Reddick2013}
Reddick R.~M., Wechsler R.~H., Tinker J.~L., Behroozi P.~S., 2013, Astrophys.
  J., 771, 30

\bibitem[{Rozo {et~al}\mbox{.}(2015)Rozo, Rykoff, Abate, Bonnett, Crocce,
  Davis, Hoyle, Leistedt, Peiris, Wechsler, Abbott, Abdalla, Banerji, Bauer,
  Benoit-L{\'{e}}vy, Bernstein, Bertin, Brooks, Buckley-Geer, Burke, Capozzi,
  Rosell, Carollo, Kind, Carretero, Castander, Childress, Cunha, D'Andrea,
  Davis, DePoy, Desai, Diehl, Dietrich, Doel, Eifler, Evrard, Neto, Flaugher,
  Fosalba, Frieman, Gaztanaga, Gerdes, Glazebrook, Gruen, Gruendl, Honscheid,
  James, Jarvis, Kim, Kuehn, Kuropatkin, Lahav, Lidman, Lima, Maia, March,
  Martini, Melchior, Miller, Miquel, Mohr, Nichol, Nord, O'Neill, Ogando,
  Plazas, Romer, Roodman, Sako, Sanchez, Santiago, Schubnell, Sevilla-Noarbe,
  Smith, Soares-Santos, Sobreira, Suchyta, Swanson, Thaler, Thomas, Uddin,
  Vikram, Walker, Wester, Zhang, \& da~Costa}]{Rozo2015}
Rozo E. {et~al.}, 2015, arXiv:1507.05460

\bibitem[{Ruiz {et~al}\mbox{.}(2015)Ruiz, Paz, Lares, Luparello, Ceccarelli, \&
  Lambas}]{Ruiz2015}
Ruiz A.~N., Paz D.~J., Lares M., Luparello H.~E., Ceccarelli L., Lambas D.~G.,
  2015, Mon. Not. R. Astron. Soc., 448, 1471

\bibitem[{Rykoff {et~al}\mbox{.}(2014)Rykoff, Rozo, Busha, Cunha, Finoguenov,
  Evrard, Hao, Koester, Leauthaud, Nord, Pierre, Reddick, Sadibekova, Sheldon,
  \& Wechsler}]{Rykoff2014}
Rykoff E.~S. {et~al.}, 2014, Astrophys. J., 785, 104

\bibitem[{Sahl{\'{e}}n, Zubeld{\'{\i}}a \& Silk(2016)Sahl{\'{e}}n,
  Zubeld{\'{\i}}a, \& Silk}]{Sahlen2016}
Sahl{\'{e}}n M., Zubeld{\'{\i}}a {\'{I}}., Silk J., 2016, Astrophys. J., 820,
  L7

\bibitem[{S{\'{a}}nchez {et~al}\mbox{.}(2014)S{\'{a}}nchez, {Carrasco Kind},
  Lin, Miquel, Abdalla, Amara, Banerji, Bonnett, Brunner, Capozzi, Carnero,
  Castander, da~Costa, Cunha, Fausti, Gerdes, Greisel, Gschwend, Hartley,
  Jouvel, Lahav, Lima, Maia, Marti, Ogando, Ostrovski, Pellegrini, Rau, Sadeh,
  Seitz, Sevilla-Noarbe, Sypniewski, de~Vicente, Abbot, Allam, Atlee,
  Bernstein, Bernstein, Buckley-Geer, Burke, Childress, Davis, DePoy, Dey,
  Desai, Diehl, Doel, Estrada, Evrard, Fernandez, Finley, Flaugher, Frieman,
  Gaztanaga, Glazebrook, Honscheid, Kim, Kuehn, Kuropatkin, Lidman, Makler,
  Marshall, Nichol, Roodman, Sanchez, Santiago, Sako, Scalzo, Smith, Swanson,
  Tarle, Thomas, Tucker, Uddin, Valdes, Walker, Yuan, \& Zuntz}]{Sanchez2014}
S{\'{a}}nchez C. {et~al.}, 2014, Mon. Not. R. Astron. Soc., 445, 1482

\bibitem[{Sheldon {et~al}\mbox{.}(2004)Sheldon, Johnston, Frieman, Scranton,
  McKay, Connolly, Budav{\'{a}}ri, Zehavi, Bahcall, Brinkmann, \&
  Fukugita}]{Sheldon2004}
Sheldon E.~S. {et~al.}, 2004, Astron. J., 127, 2544

\bibitem[{Sheth \& van~de Weygaert(2004)}]{Sheth2004}
Sheth R.~K., van~de Weygaert R., 2004, Mon. Not. R. Astron. Soc., 350, 517

\bibitem[{Song \& Lee(2009)}]{Song2009}
Song H., Lee J., 2009, Astrophys. J., 701, L25

\bibitem[{Spolyar, Sahl{\'{e}}n \& Silk(2013)Spolyar, Sahl{\'{e}}n, \&
  Silk}]{Spolyar2013}
Spolyar D., Sahl{\'{e}}n M., Silk J., 2013, Phys. Rev. Lett., 111, 241103

\bibitem[{Springel(2005)}]{Springel2005a}
Springel V., 2005, Mon. Not. R. Astron. Soc., 364, 1105

\bibitem[{Sutter {et~al}\mbox{.}(2014{\natexlab{a}})Sutter, Lavaux, Hamaus,
  Wandelt, Weinberg, \& Warren}]{Sutter2014}
Sutter P.~M., Lavaux G., Hamaus N., Wandelt B.~D., Weinberg D.~H., Warren
  M.~S., 2014{\natexlab{a}}, Mon. Not. R. Astron. Soc., 442, 462

\bibitem[{Sutter {et~al}\mbox{.}(2012)Sutter, Lavaux, Wandelt, \&
  Weinberg}]{Sutter2012}
Sutter P.~M., Lavaux G., Wandelt B.~D., Weinberg D.~H., 2012, Astrophys. J.,
  761, 44

\bibitem[{Sutter {et~al}\mbox{.}(2014{\natexlab{b}})Sutter, Pisani, Wandelt, \&
  Weinberg}]{Sutter2014b}
Sutter P.~M., Pisani A., Wandelt B.~D., Weinberg D.~H., 2014{\natexlab{b}},
  Mon. Not. R. Astron. Soc., 443, 2983

\bibitem[{Tyson {et~al}\mbox{.}(2003)Tyson, Wittman, Hennawi, \&
  Spergelb}]{Tyson:2002nh}
Tyson J.~A., Wittman D.~M., Hennawi J.~F., Spergelb D.~N., 2003, Nucl. Phys. B
  - Proc. Suppl., 124, 21

\bibitem[{Wechsler(2004)}]{Wechsler2004}
Wechsler R.~H., 2004, Clust. Galaxies Probes Cosmol. Struct. Galaxy Evol. from
  Carnegie Obs. Centen. Symp. Carnegie Obs. Astrophys. Ser. Ed. by J.S.
  Mulchaey, A. Dressler, A. Oemler, 2004. Pasadena Carnegie Ob

\bibitem[{Yang {et~al}\mbox{.}(2015)Yang, Neyrinck, Arag{\'{o}}n-Calvo, Falck,
  \& Silk}]{Yang2015}
Yang L.~F., Neyrinck M.~C., Arag{\'{o}}n-Calvo M.~A., Falck B., Silk J., 2015,
  Mon. Not. R. Astron. Soc., 451, 3606

\bibitem[{York {et~al}\mbox{.}(2000)York, Adelman, {Anderson, Jr.}, Anderson,
  Annis, Bahcall, Bakken, Barkhouser, Bastian, Berman, Boroski, Bracker,
  Briegel, Briggs, Brinkmann, Brunner, Burles, Carey, Carr, Castander, Chen,
  Colestock, Connolly, Crocker, Csabai, Czarapata, Davis, Doi, Dombeck,
  Eisenstein, Ellman, Elms, Evans, Fan, Federwitz, Fiscelli, Friedman, Frieman,
  Fukugita, Gillespie, Gunn, Gurbani, de~Haas, Haldeman, Harris, Hayes,
  Heckman, Hennessy, Hindsley, Holm, Holmgren, Huang, Hull, Husby, Ichikawa,
  Ichikawa, Ivezi{\'{c}}, Kent, Kim, Kinney, Klaene, Kleinman, Kleinman, Knapp,
  Korienek, Kron, Kunszt, Lamb, Lee, Leger, Limmongkol, Lindenmeyer, Long,
  Loomis, Loveday, Lucinio, Lupton, MacKinnon, Mannery, Mantsch, Margon,
  McGehee, McKay, Meiksin, Merelli, Monet, Munn, Narayanan, Nash, Neilsen,
  Neswold, Newberg, Nichol, Nicinski, Nonino, Okada, Okamura, Ostriker, Owen,
  Pauls, Peoples, Peterson, Petravick, Pier, Pope, Pordes, Prosapio,
  Rechenmacher, Quinn, Richards, Richmond, Rivetta, Rockosi, Ruthmansdorfer,
  Sandford, Schlegel, Schneider, Sekiguchi, Sergey, Shimasaku, Siegmund, Smee,
  Smith, Snedden, Stone, Stoughton, Strauss, Stubbs, SubbaRao, Szalay, Szapudi,
  Szokoly, Thakar, Tremonti, Tucker, Uomoto, {Vanden Berk}, Vogeley, Waddell,
  Wang, Watanabe, Weinberg, Yanny, \& Yasuda}]{York:2000gk}
York D.~G. {et~al.}, 2000, Astron. J., 120, 1579

\bibitem[{Zhao {et~al}\mbox{.}(2015)Zhao, Tao, Liang, Kitaura, \&
  Chuang}]{Zhao2015}
Zhao C., Tao C., Liang Y., Kitaura F.-S., Chuang C.-H., 2015, arXiv:1511.04299

\bibitem[{Zivick {et~al}\mbox{.}(2015)Zivick, Sutter, Wandelt, Li, \&
  Lam}]{Zivick2015}
Zivick P., Sutter P.~M., Wandelt B.~D., Li B., Lam T.~Y., 2015, Mon. Not. R.
  Astron. Soc., 451, 4215

\end{thebibliography}

\section*{Appendix}
\renewcommand{\thesubsection}{\Alph{subsection}}

\subsection{Choice of $\delta_m$}
\label{sec:appendix_new}

\textcolor{black}{The void finder presented in Sect.~\ref{sec:finder} of this paper produces a void catalog which depends on the chosen value for the maximum density contrast ($\delta_m$) of a pixel to become a void center (see Sect.~\ref{sec:algorithm}). The most significant, and hence deepest voids found by the algorithm are independent of the choice of $\delta_m$, but the total number of voids in the catalog will vary with that choice. With the fiducial value being $\delta_m = -0.30$, in this appendix we vary that value by 10\% high and low, and test the impact of these changes in the void lensing signal in the data.  }

\begin{figure}
\centering
\includegraphics[height=80mm]{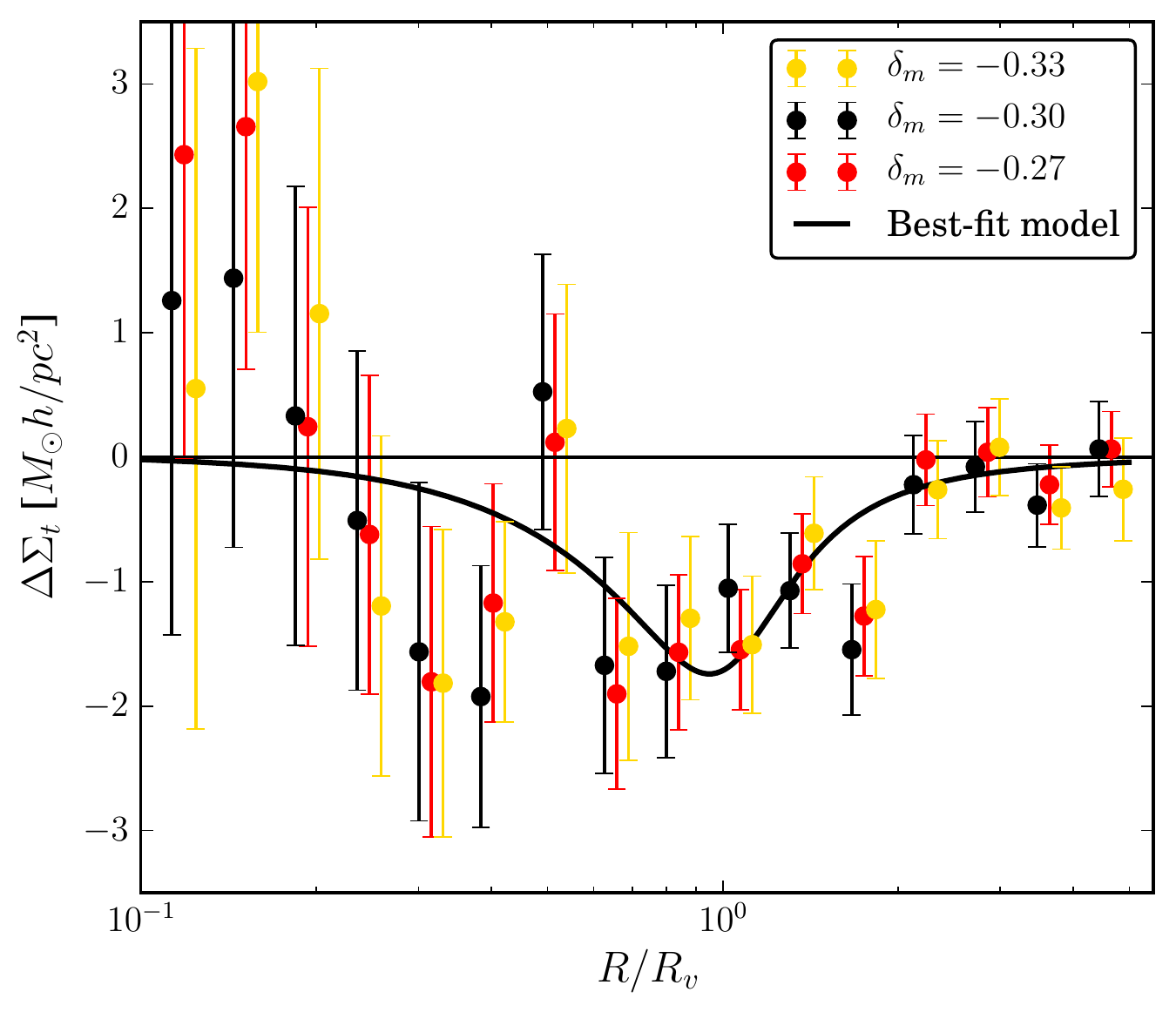}
\caption{\textcolor{black}{Stacked void lensing signal in DES-SV data for three choices of $\delta_m$: $-0.33$, $-0.30$ (fiducial), $-0.27$. The black line shows the best-fit model to the fiducial measurement. The comparison shows good agreement between the three sets of measurements.}}
\label{fig:deltam}
\end{figure}

\textcolor{black}{The fiducial void catalog with $\delta_m = -0.30$ contains 78 voids and the goodness of the best-fit model to its lensing signal (see Sect.~\ref{sec:model}) is 13.2/14. The catalog with $\delta_m = -0.33$ contains 73 voids and the goodness of the lensing fiducial best-fit model is 12.9/14. The catalog with $\delta_m = -0.27$ contains 107 voids and the goodness of the lensing fiducial best-fit model is 11.9/14. The good agreement between the lensing signal in the three cases is also shown in Fig.~\ref{fig:deltam}.} 

\subsection{Lensing on individual slicings}
\label{sec:appendix_b}

In Sect.~\ref{sec:kmeans} we presented a way of combining different slicings of the line of sight (LOS), oversampling it with slices of 100 \mpch~thickness every 20 \mpch, in order to get more information in that direction. Voids found in neighboring slices are joined if their centers are close enough, and the resulting group of voids is considered an individual physical underdensity. 

\begin{figure}
\centering
\includegraphics[height=80mm]{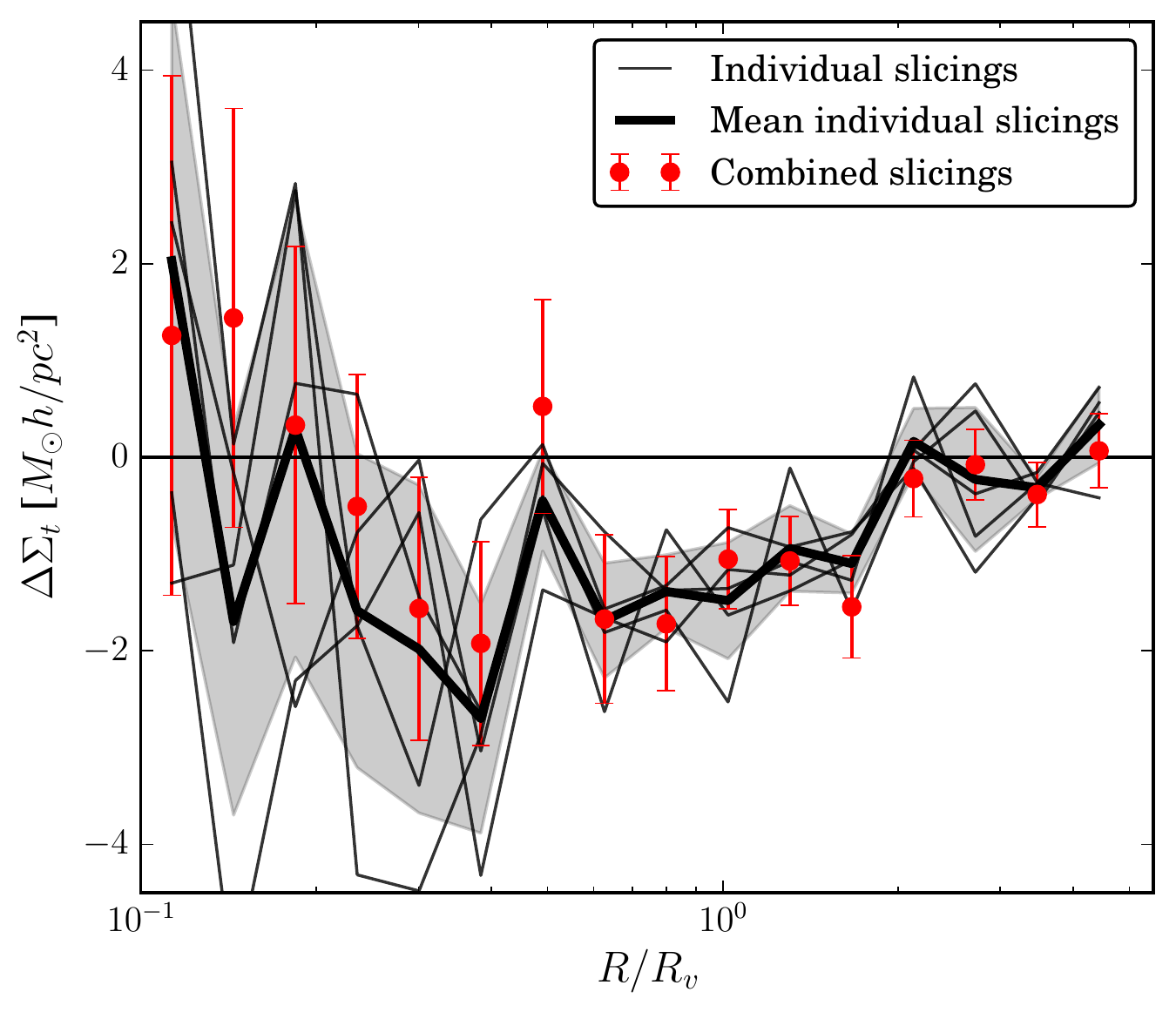}
\caption{Stacked void lensing signal in DES-SV data for each of the five individual slicings (thin black lines) and for their mean (thick black line), compared to the standard deviation of the individual slicings measurements (shaded grey region). The actual measurement of the final void catalog from Sect.~\ref{sec:lensing} is also shown (red data points with errors). This comparison shows good agreement between the combined and individual slicings.}
\label{fig:individual_slicings}
\end{figure}

In this appendix we test the impact of that procedure on the void lensing results presented in this paper (Sect.~\ref{sec:lensing}). For that purpose, we perform the lensing measurement on the set of voids found in each individual slicing, corresponding to the five columns in the graphical representation of Fig.~\ref{fig:los}. Note that in the case of individual slicings there is no overlap between the slices in which voids are found. The corresponding five lensing measurements, together with its mean and standard deviation, are shown in Fig.~\ref{fig:individual_slicings}, where they are compared to the lensing measurement presented in Sect.~\ref{sec:lensing}. The comparison in that plot, with the majority of points from the combined slicings measurement being within 1$\sigma$ of the mean individual slicings case, shows how the combined slicing approach is not affecting the lensing results in this work in any other way than reducing the noise in the measurement.  

\subsection{\textit{Randomized} void catalog}
\label{sec:appendix_a}

\begin{figure*}
\centering
\includegraphics[width=180mm]{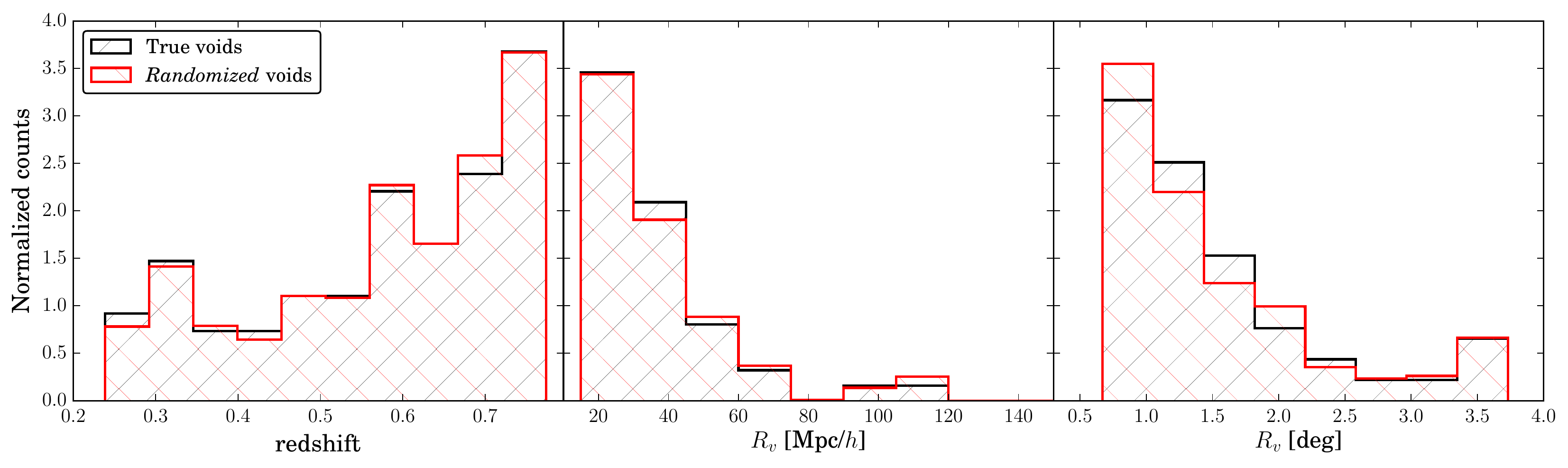}
\caption{Comparison of the true and \textit{randomized} void redshift (\textit{left panel}), comoving radius (\textit{center panel}) and angular radius distributions (\textit{right panel}). The \textit{randomized} void catalog is produced to mimic these properties of the true void catalog by following the procedure explained in Appendix \ref{sec:appendix_a}.}
\label{fig:random_voids}
\end{figure*}

The \textit{randomized} void catalog in this paper is produced such that it mimics the properties of the true void catalog in redshift and radius. We start from a set of random points inside the data mask; they will constitute the centers of the \textit{randomized} voids. We assign a redshift to each random point drawn for the true redshift distribution of voids and, to each \textit{randomized} void, we assign an angular radius from the true distribution of angular radii for voids of similar redshift (in a window of $\Delta z = 0.1$), this way preserving the redshift - angular radius relation. Finally, from the angular radius and the redshift we compute the comoving radius of the \textit{randomized} voids. 

After this process we have a \textit{randomized} void catalog with the same properties as the true one. Then, we also apply the process described in Sect.~\ref{sec:pruning} to get rid of voids near the survey edges. At the end, the \textit{randomized} void catalog has 10 times as many objects as the true one. Figure \ref{fig:random_voids} shows the agreement between the distributions of the true and \textit{randomized} voids in redshift and comoving and angular radius.

\section*{Affiliations}
$^{1}$ Institut de F\'{\i}sica d'Altes Energies (IFAE), The Barcelona Institute of Science and Technology, Campus UAB, 08193 Bellaterra (Barcelona) Spain\\
$^{2}$ Department of Physics and Astronomy, University of Pennsylvania, Philadelphia, PA 19104, USA\\
$^{3}$ Centro de Investigaciones Energ\'eticas, Medioambientales y Tecnol\'ogicas (CIEMAT), Madrid, Spain\\
$^{4}$ Institute of Cosmology \& Gravitation, University of Portsmouth, Portsmouth, PO1 3FX, UK\\
$^{5}$ Kavli Institute for Particle Astrophysics \& Cosmology, P. O. Box 2450, Stanford University, Stanford, CA 94305, USA\\
$^{6}$ SLAC National Accelerator Laboratory, Menlo Park, CA 94025, USA\\
$^{7}$ Einstein Fellow\\
$^{8}$ Universit\"ats-Sternwarte, Fakult\"at f\"ur Physik, Ludwig-Maximilians Universit\"at M\"unchen, Scheinerstr. 1, 81679 M\"unchen, Germany\\
$^{9}$ Department of Physics, University of Michigan, Ann Arbor, MI 48109, USA\\
$^{10}$ Department of Physics, ETH Zurich, Wolfgang-Pauli-Strasse 16, CH-8093 Zurich, Switzerland\\
$^{11}$ Department of Physics, Stanford University, 382 Via Pueblo Mall, Stanford, CA 94305, USA\\
$^{12}$ Department of Physics \& Astronomy, University College London, Gower Street, London, WC1E 6BT, UK\\
$^{13}$ Instituci\'o Catalana de Recerca i Estudis Avan\c{c}ats, E-08010 Barcelona, Spain\\
$^{14}$ Department of Physics, University of Arizona, Tucson, AZ 85721, USA\\
$^{15}$ Brookhaven National Laboratory, Bldg 510, Upton, NY 11973, USA\\
$^{16}$ Jodrell Bank Center for Astrophysics, School of Physics and Astronomy, University of Manchester, Oxford Road, Manchester, M13 9PL, UK\\
$^{17}$ Cerro Tololo Inter-American Observatory, National Optical Astronomy Observatory, Casilla 603, La Serena, Chile\\
$^{18}$ Department of Physics and Electronics, Rhodes University, PO Box 94, Grahamstown, 6140, South Africa\\
$^{19}$ Fermi National Accelerator Laboratory, P. O. Box 500, Batavia, IL 60510, USA\\
$^{20}$ CNRS, UMR 7095, Institut d'Astrophysique de Paris, F-75014, Paris, France\\
$^{21}$ Sorbonne Universit\'es, UPMC Univ Paris 06, UMR 7095, Institut d'Astrophysique de Paris, F-75014, Paris, France\\
$^{22}$ Carnegie Observatories, 813 Santa Barbara St., Pasadena, CA 91101, USA\\
$^{23}$ Laborat\'orio Interinstitucional de e-Astronomia - LIneA, Rua Gal. Jos\'e Cristino 77, Rio de Janeiro, RJ - 20921-400, Brazil\\
$^{24}$ Observat\'orio Nacional, Rua Gal. Jos\'e Cristino 77, Rio de Janeiro, RJ - 20921-400, Brazil\\
$^{25}$ Department of Astronomy, University of Illinois, 1002 W. Green Street, Urbana, IL 61801, USA\\
$^{26}$ National Center for Supercomputing Applications, 1205 West Clark St., Urbana, IL 61801, USA\\
$^{27}$ Institut de Ci\`encies de l'Espai, IEEC-CSIC, Campus UAB, Carrer de Can Magrans, s/n,  08193 Bellaterra, Barcelona, Spain\\
$^{28}$ School of Physics and Astronomy, University of Southampton,  Southampton, SO17 1BJ, UK\\
$^{29}$ Excellence Cluster Universe, Boltzmannstr.\ 2, 85748 Garching, Germany\\
$^{30}$ Faculty of Physics, Ludwig-Maximilians-Universit\"at, Scheinerstr. 1, 81679 Munich, Germany\\
$^{31}$ Department of Astronomy, University of Michigan, Ann Arbor, MI 48109, USA\\
$^{32}$ Kavli Institute for Cosmological Physics, University of Chicago, Chicago, IL 60637, USA\\
$^{33}$ Center for Cosmology and Astro-Particle Physics, The Ohio State University, Columbus, OH 43210, USA\\
$^{34}$ Department of Physics, The Ohio State University, Columbus, OH 43210, USA\\
$^{35}$ Australian Astronomical Observatory, North Ryde, NSW 2113, Australia\\
$^{36}$ Departamento de F\'{\i}sica Matem\'atica,  Instituto de F\'{\i}sica, Universidade de S\~ao Paulo,  CP 66318, CEP 05314-970, S\~ao Paulo, SP,  Brazil\\
$^{37}$ George P. and Cynthia Woods Mitchell Institute for Fundamental Physics and Astronomy, and Department of Physics and Astronomy, Texas A\&M University, College Station, TX 77843,  USA\\
$^{38}$ Department of Astrophysical Sciences, Princeton University, Peyton Hall, Princeton, NJ 08544, USA\\
$^{39}$ Jet Propulsion Laboratory, California Institute of Technology, 4800 Oak Grove Dr., Pasadena, CA 91109, USA\\
$^{40}$ Department of Physics and Astronomy, Pevensey Building, University of Sussex, Brighton, BN1 9QH, UK\\
$^{41}$ ICTP South American Institute for Fundamental Research\\ Instituto de F\'{\i}sica Te\'orica, Universidade Estadual Paulista, S\~ao Paulo, Brazil\\
$^{42}$ Computer Science and Mathematics Division, Oak Ridge National Laboratory, Oak Ridge, TN 37831, USA\\
$^{43}$ Max Planck Institute for Extraterrestrial Physics, Giessenbachstrasse, 85748 Garching, Germany\\

\bsp
\label{lastpage}

\end{document}